\documentclass[aps,prc,twocolumn,groupedaddress]{revtex4-1}

\usepackage{graphicx}
\usepackage{amsmath}
\usepackage{xcolor} 

\begin{document}


\title{Coupling between collective modes in the deformed $^{98}$Zr nucleus: Insights from consistent HFB+QRPA calculations with the Gogny interaction}


\author{E. V. Chimanski$^{1}$}
\email{Current address at Brookhaven National Laboratory: chimanski@bnl.gov}
\author{E. J. In$^{1}$}
\author{S. P{\'e}ru$^{2}$}%
\author{A. Thapa$^{1}$}
\author{W. Younes$^{1}$}
\author{J. E. Escher$^{1}$}


\affiliation{%
 Lawrence Livermore National Laboratory, Livermore, CA, USA $^{1}$ \\
CEA, DAM, DIF, F-$91297$ Arpajon, France $^{2}$ \\
Université Paris-Saclay, CEA, LMCE, $91680$ Bruyères-le-Châtel, France $^{2}$\\
}%

\date{}

\begin{abstract}
The Zirconium isotopes exhibit structural properties that present multiple challenges to nuclear theory. Investigations of the coupling present within isoscalar modes and within isovector modes are scarce but important for advancing our understanding of the microscopic picture of nuclei. To explore some of these underlying coupling features, and to test the predictive power of a state-of-the-art nuclear structure approach, we provide a detailed analysis of the properties of $^{90,96,98}$Zr. This region includes a benchmarking case and offers insights into nuclear deformation phenomena. To investigate the coupling between collective modes in deformed nuclei, we focused our analysis on the ground and excited-state properties of these isotopes, employing a consistent approach with the axially-symmetric deformed Hartree-Fock-Bogoliubov (HFB) and the Quasiparticle Random Phase Approximation (QRPA) framework, both using the Gogny D1M force.  This approach effectively describes both low-lying and giant-resonance states. We devoted special attention to the deformed $^{98}$Zr nucleus, where we confirm the existence of coupling between monopole and quadrupole excitations through the $K^{\pi} = 0^{+}$ QRPA components and demonstrate an analogous dipole-octupole coupling through the $K^{\pi} = 0^{-}$ and $K^{\pi} = 1^{-}$ components. Intrinsic transition densities and associated radial projections illustrate the coupling. Our work complements and extends earlier studies  carried out using density-functional-based methods  and notably, we included the complete Coulomb interaction also in the pairing fields, i.e. we treat terms exactly that are approximated in typical calculations that use the Gogny D1 and D2 interaction families.
\end{abstract}


\maketitle

\section{Introduction}
\label{intro}
Microscopic descriptions of nuclei are key to understanding and predicting static and dynamic properties of nuclei. They provide insights that complement and greatly enhance phenomenological approaches, such as the semiclassical models based on a liquid-drop concept~\cite{Mottelson-97}.  Methods based on density-functional theory (DFT) such as mean-field theory (MFT), which are the starting point of the present work, provide a microscopic many-body framework that accommodates realistic effective nuclear interactions, pairing fields and quasi-particle (qp) wave functions~\cite{Hergert:20,Ring-2004,Suhonen-2017}. 
These methods allow us to calculate ground state properties, such as nuclear binding energies, particle separation energies, and charge radii.  Careful comparisons with existing data is crucial for assessing the predictive power of the chosen model and can help to improve the framework or interaction used.  A properly benchmarked model can then be used to investigate the microscopic structures that are associated with these observables and to predict unknown properties of nuclei and trends for isotopes far from stability.
Of particular interest is understanding the evolution of shell structure, the emergence of deformation, and collective excitations~\cite{Federman-1979,Goeke-1982,Heyde-2011,Garg-2018,Carlson-2020}. 
Beyond-mean-field theories, such as the Quasiparticle Random Phase Approximation (QRPA) employed in the present study, can also predict properties of excited states.  The QRPA approach is particularly well suited to study low-lying vibrational states and giant resonance excitations for both parity of states in the same framework \cite{Ring-2004,Utsunomiya-2008,Peru-2014,Martini-2016,Deloncle-2017,Goriely-2018,Nakatsukasa-2016,Goriely:19}.  
Describing and understanding the emergence of collective excitations from the underlying microscopic structures is an important goal of nuclear physics.

In the present paper, we will focus on the coupling between collective nuclear excitations with a significant implications for understanding nuclear structure and reaction dynamics. The monopole-quadrupole coupling is an effect well known from studies of low-lying beta vibrations \cite{Goeke-1982,Bahini-2022} playing a key role in determining the compressibility of nuclear matter and the nuclear equation of state (EoS), which affects our understanding of neutron stars \cite{Blaizot-1980,haensel/2007,Lattimer-2012,RocaMaza-2018,Xuwei2022,Adri2023}.  The coupling between dipole and quadrupole vibrations  in Sn isotopes were explored by Simenel and Chomaz \cite{Simenel-2009} using a time-dependent Hartree-Fock methods based on the Skyrme functional. These coupling affect the nuclear motion introducing  further complexity into the multi-phonon spectrum.  Yoshida and Nakatsukasa \cite{Yoshida-2013} performed QRPA calculations using the Skyrme-energy-density-functional method to demonstrate that deformation enables mixing between dipole and octupole excitations and leads to broadening of the Isoscalar Giant Dipole Resonance (ISGDR) and enhancement of low-lying dipole strengths. As the details of this coupling remain to be understood, we further explore and illustrate it in depth in this work. The ability to predict these resonances can have a broader impact given that the dipole response influences low-energy neutron capture reactions, which are important for astrophysics simulations~\cite{Arcones-17} and applied nuclear physics~\cite{Capote-2009,Goriely:19}.
The structure predictions considered in this paper play a more general role in nuclear reaction calculations: The isovector dipole resonance is related to the $\gamma$-ray strength function, which is an important input for statistical (Hauser-Feshbach) reaction calculations~\cite{Hauser-1952,Capote-2009}. Moreover, the transition densities extracted from the QRPA calculations can be folded with an effective projectile-nucleon interaction to produce coupling potentials for distorted-wave Born Approximation (DWBA) or coupled-channels (CC) direct-reaction calculations~\cite{Nobre-PRL-2010, Nobre-PRC-2011, Dupuis-PRC-2019, Thapa:2024gyy}.  Thus, they represent a step towards a predictive direct-reaction theory.
Such predictive capabilities will play an important role in interpreting data from new radioactive beam facilities~\cite{Balantekin-2014, Johnson-2020, Hilaire2021}.  In turn, data for exotic isotopes far from stability will challenge the predictive power of theory and provide important touchstones for future developments.  

The present work focuses on a selected set of Zr isotopes, specifically $^{90,96,98}$Zr.  The $^{90}$Zr nucleus is very well studied, and provides good opportunities to compare our calculations to experimental data or other calculations.  At the same time, the Zr isotopes exhibit a rich set of phenomena and open questions, in particular in the areas of shape evolution and shape coexistence.  Our work complements existing theoretical studies on the structure of Zr chain of isotopes, many of which focused on a set of isotopes or used semi-phenomenological approaches, such as Ref.~\cite{Ramos-2019}, which invokes the interacting boson model (IBM). Other approaches include  Relativistic Hartree-Bogoliubov (RHB) calculations with the density dependent meson exchange model (DDME2) \cite{Thakur-2021}, Hartree-Fock (HF) and Hartree-Fock-Bogoliubov (HFB) studies with Skyrme and M3Y-P6 interactions~\cite{Blazkiewicz-2005, Miyahara-2018}.  The QRPA study presented in Ref.~ \cite{Deloncle-2017} uses the D1M Gogny interaction  in an approach very similar to ours to study the dipole response, but considerations were limited to the spherical $^{90-94}$Zr isotopes.  

Specifically, we provide a detailed description of ground-state properties and shape evolution in the even-even $^{90,96,98}$Zr isotopes, obtained from HFB+QRPA calculations in an axially-symmetric deformed basis, using the D1M Gogny effective interaction self-consistently for both ground and excited-state calculations. As benchmark, the low-lying collective states and B(E2) values are compared to experimental data for the spherical cases. Mean energy values for the giant monopole, dipole, and quadrupole resonances are calculated and compared to available experimental values and systematic formulae. We specifically discuss in detail the fragmentation of the electromagnetic contributions into different $K^{\pi}$ components for deformed systems and most notably, we explore the isoscalar monopole-quadrupole and isovector dipole-octupole couplings in the deformed $^{98}$Zr nucleus, providing new insights into the coupling between collective modes in deformed nuclei.

This paper is organized as follows: Sec.~\ref{HFB_results} provides a brief summary of our Hartree-Fock-Bogoliubov ground state model and the respective discussion of the obtained results (extra details can be found in the Appendix \ref{HFB_implementation_and_numerical_convergence}). Section \ref{QRPA} contains a short description of our QRPA calculations along with validation cases for the spherical nuclei. In Section ~\ref{Results}, we provide a detailed analysis of excited states including the coupling of monopole-quadrupole and dipole-octupole responses. Sec.~\ref{summary} contains a summary of our results together with an outlook for future work in the context of microscopic mean field models.  Appendix \ref{HFB_implementation_and_numerical_convergence} contains some aspects of our HFB implementation and a summary table of the obtained ground state properties. We provide in appendixes \ref{Angular_momentum_restoration}-\ref{Symmetry_considerations} the description of the angular-momentum restoration techniques used and details on how we utilized the symmetries in our basis to construct the full three-dimensional transition densities from a calculation restricted to a subspace.

 
 \section{Theory}
 
 \subsection{Ground state model}
  \label{HFB_results}

We predict nuclear ground states properties within the Hartree-Fock-Bogoliubov (HFB) framework, implemented in a cylindrical harmonic oscillator basis (HO). Our calculations preserve axial and reflection symmetry.  All our calculations are performed in bases that span 11 major oscillator shells (i.e., with a maximum shell number $N_{\rm{osc}} =10$).
For each isotope, we explored a wide range of values for the deformation parameter $\beta$ which serves as a constraint in the HFB calculations; it is defined through:
\begin{equation}
    \beta = \sqrt{\frac{5}{9}\pi}\frac{q_{20}}{AR^{2}},
\end{equation}
where $q_{20}$ denotes the mean value of the axial quadrupole operator, $A$ is the number of nucleons, and $R=1.2 A^{\frac{1}{3}}$ fm is the nuclear radius.
The oscillator lengths are adjusted individually for each isotope and kept equal ($b_{0}=b_{z}=b_{\perp}$), a prescription allowed by the large basis size. 
The relationship between the calculated energy $E$ and the deformation $\beta$ defines an energy curve for each isotope. For the $^{90}$Zr case, we performed tests with smaller and larger bases to confirm the convergence of the numerical results (details of our HFB implementation as well as the convergence tests are described in the Appendix \ref{HFB_implementation_and_numerical_convergence}). 
Our HFB calculations are performed with the finite-range Gogny force~\cite{Decharge-1980,Berger-1984,Berger-1991,Goriely-2009} and the Coulomb interaction. The Gogny interaction has proven to be very successful in globally describing the nuclear structure properties in the mean field approach. The successful predictive power of the interaction goes beyond the mean-field level, and we refer to \cite{Robledo-2019} for a more detailed exploration of this aspect.
 A handful of parameterizations exists, each of which builds on a predecessor and is obtained by including technical improvements and new measurements in the fitting process. Here we choose the D1M parameterization \cite{Goriely-2009}, which achieves better performance in describing nuclear masses and nuclear/neutron matter properties including quadrupole correlations. For selected cases, we also performed comparisons to results obtained with the earlier D1S parameterization; these are explicitly indicated.

The parameterization of the D1 family was obtained in a fitting process that did not include the Coulomb exchange contribution during the iterative steps of the HFB calculations. Here, however, we were able to leverage our framework to investigate the impact of using the exact treatment for the Coulomb terms by comparing to results obtained using the earlier approximation. We focus on the Coulomb contributions because these were found to be particularly important when going beyond standard approximations~\cite{Anguiano-2001,Anguiano-NPA-2001}. Effects due to the inclusion of the spin-orbit interaction and two-body center-of-mass corrections in the pairing field are expected to be smaller and are not considered in the present study.

\subsubsection{Predicted ground state properties of the Zirconium isotopes and the impact of the interaction on ground state energies}

\begin{figure}[htb]
\centering
\includegraphics[scale=0.57]{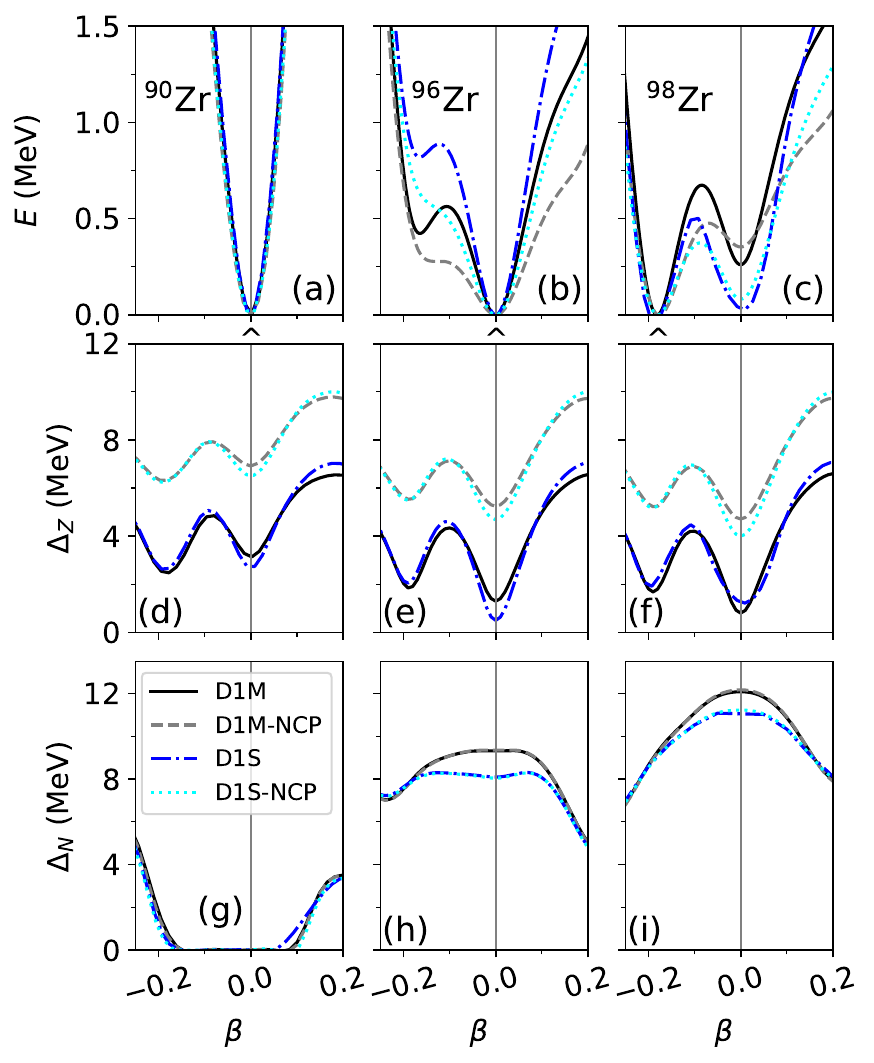}
\caption{\label{Zr_comp}  Numerical results with the D1M and D1S Gogny interactions, as well as different treatments of the Coulomb contribution for $^{90,96,98}$Zr isotopes .  The top, middle, and bottom panels show the total HFB energy curve, and the energy of proton and neutron pairing fields, respectively. Calculations with the exact treatment of the Coulomb fields in the D1M and D1S are indicated by solid black (D1M) and dotted-dashed blue (D1S) lines, respectively. Calculations resulting from a treatment that omits the Coulomb contributions to the pairing fields are indicated with dashed gray (D1M-NCP) and dotted cyan lines (D1S-NCP). The marks represent the ground state shape and energy  where the QRPA excitations were computed for each isotope.}
\end{figure}

We systematically constructed energy curves by performing HFB calculations for various quadrupole deformations. The minimum in this curve determines the HFB ground state energy $E_{\textrm{HFB}}$ and deformation $\beta_{\textrm{HFB}}$, where a $\beta_{\textrm{HFB}} = 0$, $<0$, and $>0$ are interpreted as representing spherical, oblate and prolate ground state shapes, respectively. The HFB energy curves for the $^{90,96,98}$Zr isotopes, shown in Figure \ref{Zr_comp}, highlight the predicted ground state shape within our axially symmetric framework  and the influence of the interaction parameterization on ground state energies. $^{90}$Zr and $^{96}$Zr are spherical in their ground states, however, $^{96}$Zr exhibits a more complex energy curve as one approaches the deformed $^{98}$Zr isotope. Shape coexistence has been observed for $^{98}$Zr \cite{Wu-2004,Guzman-2010,Heyde-2011,Singh-2018,Witt-2018} and for $^{94-96}$Zr \cite{Guzman-2010,Chakraborty-2013,Kremer-2016} reflecting a complex transition region for the Zr isotopes. Different theoretical studies have also reported local minima in this region of the isotopic chart \cite{Thakur-2021,Miyahara-2018}. 5DCH model calculations in a triaxial basis \cite{amedee_database} support our global minimum solutions. Detailed Generator Coordinate Method (GCM) calculations \cite{Hill-1953, Griffin-1957,Guzman-2010,Rodriguez-2011, Younes-2019} would be required to study possible mixing of different HFB solutions and shed further light on shape phenomena.

It is informative to study the impact of approximations to the treatment of the Coulomb fields commonly employed in such calculations. This was motived by the fact that both the D1M and D1S parameterizations were obtained in a fitting process that ignored the exchange contributions from the Coulomb interaction in the HFB iterations for all but the last step, where the contribution from the Slater approximation is added to the final energy. This is different from the exact treatment that we employ throughout this paper.  To assess the impact of the Coulomb exchange in the pairing terms, we performed select D1M and D1S calculations without this contribution (denoted as `NCP' for `No-Coulomb-in-Pairing')  and compared to our exact treatment. The differences observed do not affect the general conclusions of our work, but illustrate the sensitivity of exotic nuclear shapes to changes in the interaction.  Our findings also reinforce the value of developing computational frameworks that fully accommodate all terms in the interaction and the need to employ such tools in future efforts to update the Gogny interaction.  The effect is particularly pronounced for the proton pairing energy, as can be seen in the middle panel of Fig.~\ref{Zr_comp}. The energy of the neutron pairing fields are barely affected in this case (bottom panel of Fig.~\ref{Zr_comp}). The impact of Coulomb anti-pairing observed is in line with recent extensive studies of pairing effects in fission and for a wide range of heavy nuclei \cite{Rod2022, Rod2023}.

\subsection{Quasiparticle Random Phase Approximation for excited states}
\label{QRPA}
We describe excited nuclear states in the Quasiparticle Random Phase Approximation (QRPA) represented in a matrix form as
 \begin{equation}
 \label{QRPA_matrix}
\begin{bmatrix}
A           & B  \\
B ^{*} & A^{*}
\end{bmatrix}
\begin{bmatrix}
X \\
Y
\end{bmatrix}
=
\omega 
\begin{bmatrix}
I & 0 \\
0 &-I
\end{bmatrix}
\begin{bmatrix}
X \\
Y
\end{bmatrix}.
 \end{equation}
The sub-matrices represent matrix elements of the nuclear interaction that include rearrangement terms and proper phase factors, see e.g. Refs.~\cite{Ring-2004,Suhonen-2017,Peru-2014} for more details. Here, we also employ the Gogny force, and to ensure consistency of our calculations, we use the same parametrization at both the HFB and QRPA levels, and we treat the Coulomb term in the same way.
Diagonalization of Eq.~(\ref{QRPA_matrix}) gives the QRPA energy spectrum and the wave functions of the excited states. The axially symmetric QRPA problem is solved for a given value of the projection of angular momentum and parity $K^{\pi}$, where the quasiparticle states (Latin letters $i$ and $j$) are combined to ensure that $K = k_{i} +k_{j}$ and $\pi = \pi _{i}\,\pi _{j}$.

The QRPA phonon creation operators can be constructed from the $X$ and $Y$ components:
\begin{eqnarray}
\label{QRPA_creation}
\hat{\theta }^{\dagger}_{n,K} = \sum _{i<j}\left ( X^{ij}_{n,K}\eta ^{\dagger}_{k_{i}}\eta ^{\dagger}_{k_{j}} + Y^{ij}_{n,K}\eta _{-k_{j}}\eta _{-k_{i}}\right ),
\end{eqnarray}
and QRPA excited states $|\theta ^{\dagger}_{n},K \rangle$ are obtained by acting with phonon operators $\hat{\theta }^{\dagger}_{n,K}$ on the ground state of the even-even nucleus under consideration:
\begin{eqnarray}
\label{QRPA_gsCondition}
  |\theta ^{\dagger}_{n},K \rangle = \hat{\theta }^{\dagger}_{n,K}|0_{\text{def}},(K=0)\rangle .
  \end{eqnarray}
Time-reversed states are represented by $|\theta ^{\dagger}_{\bar{n}},-K \rangle$. The subscript $n$ denotes the n-th eigenstate of the QRPA spectrum for given $K^{\pi}$ and $\bar{n}$ labels the time-reversed eigenstate associated with the $-K^{\pi}$ state at the same energy. The exact ground state is defined by the condition $ \hat{\theta }_{n,K} |0_{\text{def}},(K=0)\rangle = 0$, i.e. it is destroyed by the action of any QRPA destruction operator.
\newpage
\subsubsection{QRPA response for spherical and deformed nuclei}
\label{QRPA_Response}

Here, we restrict our considerations to axially-symmetric, deformed nuclei, i.e. the $K$ quantum number, which gives the projection of the angular momentum onto the symmetry axis, is preserved. To obtain the full response for an operator with given angular-momentum and parity $J^{\pi}$, we need to calculate contributions from multiple $K^{\pi}$ components.  More specifically, we have to solve the QRPA problem for $K^{\pi}=0 ^{\pi},\pm 1^{\pi},\ldots, \pm J^{\pi}$. In the limit of spherical symmetry, the different $K^{\pi}$ components are degenerate in energy, and one can extract the response for any multipolarity directly from $K^{\pi} = 0 ^{\pm}$ states (see for example \cite{Martini-2016,chimanski-2022}).

To obtain the multiple components of the QRPA response for an operator with quantum numbers $\lambda \mu$ (denoting angular momentum and projection on the 3-axis), we use standard angular-momentum restoration techniques. The resulting expression involves matrix elements expressed in the intrinsic (body-fixed) coordinate system and geometric weighing factors. After some algebra (see Appendix \ref{Angular_momentum_restoration}) ones obtains for transitions from the ground to excited states
\begin{widetext}
\begin{eqnarray}
\label{projected_response}
  \langle JMK_{n} |\hat{Q}_{\lambda \mu}|\tilde{O}_{(J^{\pi}=0^{+})} \rangle = \sqrt{2J+1} \sum _{\mu ^{\prime}}(-1)^{\mu -\mu ^{\prime}} 
 \begin{pmatrix}
 J  & \lambda & 0\\
  M  &  -\mu & 0 
  \end{pmatrix}
\begin{pmatrix}
 J & \lambda &0 \\
 K & -\mu ^{\prime}& 0 
  \end{pmatrix}
  \langle \theta _{n},K |\hat{Q}_{\lambda \mu ^{\prime}}|0_{\text{def}} \rangle ,
\end{eqnarray}
\end{widetext}
where for the even-even nuclei considered here, the 3-j symbols on the r.h.s of the Eq.~(\ref{projected_response}) ensure that non-zero matrix elements occur only for $J=\lambda$ and $K= \mu ^{\prime}$.

The operator $\hat{Q}_{\lambda \mu} = r^{\lambda }Y_{\lambda \mu}$ is used to obtain the reduced electromagnetic transition probabilities, B(EJK) \footnote{Center of mass correction is taken into account for the electric dipole case and $r^{2}$ is employed for monopole response calculations.}, that can be obtained by setting $\mu = M =0$ and squaring the amplitudes calculated with Eq.~(\ref{projected_response}). Here, the contributions from the negative $K$ values can be obtained as follows:  
\begin{eqnarray}
\left | \langle  JM-K_{n} |\hat{Q}_{\lambda -\mu}|\tilde{O}_{(J^{\pi}=0^{+})}\rangle  \right |^{2} = \nonumber\\ \left | 
\langle JMK_{n}|\hat{Q}_{\lambda \mu}|\tilde{O}_{(J^{\pi}=0^{+})}  \rangle \right |^{2} .
\end{eqnarray}

We are also interested in calculating the QRPA radial transition densities, which in addition to helping us visualize the nuclear shape oscillations, play an essential role in generating transition potentials for inelastic scattering and charge-exchange cross-section calculations \cite{Nobre-PRC-2011,Nobre-PRL-2010,Dupuis-PRC-2019,Hilaire2021,Thapa:2024gyy}. These radial transitions can be obtained in a multipole expansion
 \begin{equation}
 \label{rho_LSJ}
    \rho ^{n,K}_{J} (r) = \int d\Omega \, \rho ^{n,K}(\vec{r})\, Y_{JK}(\Omega),
\end{equation}
where $Y_{JK}(\Omega)$ are the spherical harmonics and $\rho ^{n,K}(\vec{r})$ the intrinsic transition density obtained in the cylindrical coordinate system (defined in the next section).

\subsubsection{QRPA intrinsic transition densities}

Matrix elements of one-body operators are conveniently obtained in second quantization, as this formalism allows us to separate the action of the operator from the structure of the many-body states.
For example, the transition density is the matrix element of the one-body density operator $\hat{\rho} \rightarrow c^{\dagger}_{\alpha} c _{\beta}$ between initial and final nuclear states. In the intrinsic frame, the density for a transition from the ground state to an excited QRPA state is given by
\begin{equation}
  \label{td}
\rho ^{n,K}(\vec{r})= \sum _{\alpha \beta } \phi ^{*}_{\alpha}(\vec{r} ) \phi _{\beta}(\vec{r}) \, \langle \hat{\theta } _{n},K|c^{\dagger}_{\alpha }c_{\beta} |\tilde{0} \rangle ,
\end{equation}
where $\phi _{\beta}(\vec{r})$ are the spatial single particle wave functions in the cylindrical harmonic oscillator basis. The translation of (\ref{td}) into Cartesian coordinates can be performed following the symmetries considerations detailed in Appendix \ref{Angular_momentum_restoration}. The matrix element in the r.h.s of Eq.~(\ref{td}) carries the information provided by the nuclear structure model. It is also known as the spectroscopic amplitude
\begin{equation}
  \label{ZabnK}
Z^{n,K}_{\alpha,\beta }\equiv  \langle \hat{\theta } _{n},K|c^{\dagger}_{\alpha }c_{\beta} |\tilde{0}\rangle .
\end{equation}
It is evaluated using the QRPA operator (Eq.~\ref{QRPA_creation}), the inverted Bogoliubov transformations 
\begin{eqnarray}
c_{\alpha}^{\dagger}= \sum _{i}\left ( U^{*}_{\alpha i} \, \eta ^{\dagger}_{i}+V_{\alpha i} \, \eta _{i} \right ), \quad c_{\alpha}= \sum _{i}\left ( V^{*}_{\alpha i} \, \eta ^{\dagger}_{i}+U_{\alpha i}\, \eta _{i}\right ) \nonumber,
\end{eqnarray}
the Quasi-Boson Approximation (QBA) and the fact that \cite{Ring-2004, Suhonen-2017}
\begin{equation}
\langle \text{HFB}| \eta _{i^{\prime}}\eta _{j^{\prime}}\eta ^{\dagger}_{i}\eta ^{\dagger} _{j}|\text{HFB}\rangle = \delta _{i^{\prime}j}\delta _{j^{\prime}i}-\delta _{i^{\prime}i}\delta _{j^{\prime}j}.
\end{equation}
Explicitly, the matrix elements take the following form in the quasi-particle basis
\begin{eqnarray}
\label{Z_ab}
Z^{n,K}_{\alpha,\beta } =
  &&  \, \sum _{i<j} \bigg [ X^{ij}_{n,K} \left ( U_{\alpha i} V_{\beta j} - U_{\alpha j} V_{\beta i} \right ) \nonumber \\
    &&\qquad +  Y^{ij}_{n,K} \left ( V_{\alpha j}U_{\beta i} -  V_{\alpha i}U_{\beta j} \right )  ]
\end{eqnarray}
 Here, $X^{*} = X$, $Y^{*}=Y$, $U=U^{*}$ and $V^{*}=V$, since the matrix elements are real-valued. Similarly, the matrix elements of a general one-body operator $\hat{Q}_{\lambda \mu}$ can be expressed in terms of the spectroscopic amplitudes:
\[
\langle \hat{\theta } _{n},K| \hat{Q}_{\lambda \mu} |\tilde{0}\rangle = \sum_{\alpha \beta} \langle \alpha| \hat{Q}_{\lambda \mu} |\beta \rangle  \, Z^{n,K}_{\alpha,\beta },
\]
where $\langle \alpha| \hat{Q}_{\lambda \mu} |\beta \rangle$ are the matrix elements of $ \hat{Q}_{\lambda \mu}$ in the single particle basis $\phi _{\alpha}$.

\subsubsection{Treatment of spurious states}

Before discussing our QRPA results, we comment on our process for removing spurious states and restoring symmetries in the QRPA calculations presented. The topic of spurious states has been extensively discussed in the literature \cite{Thouless-1961,Bender-2003,Ring-2004,Colo-2013,Peru-2014,Repko-2019}. In order to briefly describe the procedure we employ in our calculations, we list in Table~\ref{tab:surious_QRPA} the number of expected spurious states for our calculations.
We note that the spurious $K^{\pi} = 0^{+}$ states associated with particle number conservation appear only for finite pairing values for the respective (proton or neutron) components.
For example, a nucleus with $\Delta _{N} =0$ but $\Delta _{Z} \neq 0$ will lead to one spurious state associated with proton number non-conservation.
Spurious $K^{\pi} = \pm 1^{+}$ states associated with spurious rotational motion are only present for deformed nuclei, since spherical nuclei do not break rotational symmetry.

\begin{table}[ht]
\caption{\label{tab:surious_QRPA}Broken symmetry and expected number of spurious states in the QRPA spectrum for given $K^{\pi}$.}
\begin{center}
\begin{tabular}{c | c | c} 
\hline
 $K^{\pi}$          & Symmetry & $\#$ of spurious states  \\
\hline
\hline
 $0^{-}$ & translational & 1\\
 $0^{+}$ & particle number & 2 (one for each isospin finite pairing) \\
 $|1|^{-}$ &translational & 1\\
 $|1|^{+}$  & rotational & 1 (for deformed nuclei) \\
\hline
\end{tabular}
\end{center}
\end{table}
We follow standard procedures for identifying and eliminating spurious states in our calculations.  Most importantly, we take steps to obtain the spurious states at energies as low as possible.
We perform consistent calculations, i.e. we employ the same interaction and approximations at both HFB and QRPA levels. We use large model spaces and test the convergence of our HFB+QRPA calculations. We include all available 2qp configurations in the valence space for a given $K^{\pi}$ and do not use cut-off strategies for single-particle excitations. In this way, the spurious states are expected to be close to zero energy, allowing us to separate them from the physical spectrum \cite{Peru-2014,Martini-2016}.

\subsubsection{Validation and limitations of the QRPA calculations}
\label{QRPA_results}

Here we present QRPA results for three Zr isotopes, $^{90}$Zr, $^{96}$Zr, and $^{98}$Zr. We have performed HFB+QRPA calculations consistently by employing the same interaction and size of the harmonic oscillator basis. We analyzed $^{90}$Zr, a nucleus that has been extensively studied both experimentally and theoretically. The nuclei $^{96}$Zr and $^{98}$Zr were chosen since they are involved in the spherical-deformed transition along the chain of Zr isotopes.  
By comparing the predictions for these two isotopes we can study in detail the effect of deformation. In what follows, the QRPA energy spectra and electromagnetic responses for the selected isotopes will be presented and compared to available experimental data.

In Table \ref{tab:9096Zr_Es} we list the energies of the first $K^{\pi}=J^{\pi}=0^{+},2^{+},4^{+}$ and $3^{-}$ QRPA excited states for the spherical $^{90,96}$Zr isotopes. Throughout the text, we will use the subscript notation $K^{\pi}_{n}$ in reference to the n-th QRPA excited state. The QRPA calculations predicts the first $ 0^{+}_{1}$ excited energies with good accuracy for both isotopes. The energy of the $2^{+}_{1}$ and $4^{+}_{1}$ states are also well reproduced by our calculations. Experimentally, the first $4^+$ state of $^{96}$Zr may have an interesting structure, as there is no reported E2 transition to the $2^+_1$ state, while another $4^+$ state at a slightly different energy exhibits strong E2(+M3) transitions to lower-energy $2^+$ states ~\cite{Abriola-2008,ENSDF_url}. In the latter case, the presence of transitions to a $3^-$ state and another $2^+$ state suggests the existence of single-particle or particle-hole components.  This argument could be supported by the fact that the pairing gap for $^{96}$Zr is $\sim1.4$ MeV, approximately half of the 2.8 MeV excitation energy of the state, is potentially indicating a mixing between single-particle excitations.  Our QRPA calculations describe the $4^+_1$ state as a QRPA excitation, but a more detailed theoretical treatment incorporating transitions between excited states would be necessary to gain further insight into the nature of the $4^+_1$ state and improve comparisons for this case. The energy of the first $3^{-}$ state agrees well with experiment for the $^{90}$Zr nucleus while for the $^{96}$Zr case, the QRPA predicts a smaller value. We note that the inclusion of Coulomb contribution in the pairing fields, i.e. the complete treatment of the Coulomb interaction, is responsible for lowering the energy of this first excited state. We found that a consistent HFB+QRPA calculation without the complete treatment this state is at 1.605 MeV, giving a closer value to the experimental data in this case. Here, we are able to incorporate the exact Coulomb exchange in both the mean field and pairing terms throughout the iterative HFB process and the QRPA calculations. This more complete computational framework allows us to highlight areas where future improvements in the force may impact theoretical predictions. These low-energy states often correspond to collective vibrations of the nucleus.  The possibility of weakly rotational bands in the deformed $^{98}$Zr nucleus together with the presence of multiple low-lying bandheads with irregular level spacing in evaluated levels in ENSDF, makes a direct comparison between our QRPA results and experimental energies  difficult. The complexity of the level scheme suggests that a more detailed experimental and theoretical evaluation is needed to clarify the nature of these excitations. Without introducing a model for rotational excitations~\cite{Bertulani-2007,Rowe-2010}, our QRPA calculations give the energies of the bandheads only. Introducing rotational degrees of freedom and calculating moments of inertia and associated excitations will be addressed in future work.
\begin{table}[ht]
\caption{\label{tab:9096Zr_Es}QRPA energies for first excited $K^{\pi}=J^{\pi}=0^{+},2^{+},4^{+}$ and $K^{\pi}=3^{-}$ states for the spherical $^{90}$Zr and $^{96}$Zr isotopes.  Experimental values are taken from \cite{Nudat}.
$^{*}$Better agreement with experiment is obtained when a calculation similar to the conventional procedure for treating the Coulomb contribution is performed ($E(3^{-}_{1}) = 1.605$ MeV, see text). }
\begin{center}
\begin{tabular}{c c c c} 
\hline
Isotope &$J^{\pi}$ & QRPA (MeV)  & Exp (MeV)\\
\hline
\hline
$^{90}$Zr&$0^{+}_{1}$ & 1.658  & 1.760 \\
$^{90}$Zr&$2^{+}_{1}$ & 2.725  & 2.186 \\
$^{90}$Zr&$4^{+}_{1}$ & 3.215  & 3.076 \\
$^{90}$Zr&$3^{-}_{1}$ & 2.858  & 2.748 \\
\hline
$^{96}$Zr&$0^{+}_{1}$ & 1.054  & 1.581 \\
$^{96}$Zr&$2^{+}_{1}$ & 1.815 & 1.750 \\
$^{96}$Zr&$4^{+}_{1}$ & 2.740 & 2.750 \\
$^{96}$Zr&$3^{-}_{1}$ & 0.553 $^{*}$& 1.897 \\
\hline
\end{tabular}
\end{center}
\end{table}

An important strength of the QRPA approach lies in its ability to predict both low-lying collective states and giant resonances (GR).
Empirical formulae have been developed for the average excitation energies of three giant resonance modes: the isoscalar monopole (ISGMR), the isovector dipole (IVGDR) and the isoscalar quadrupole (ISGQR) resonances.  Their energies can be estimated using simple formulae that depend only on the number of nucleons $A$ in the nucleus \cite{Harakeh-2001,Martini-2016}:
\begin{eqnarray}
\label{E_GR_syst}
E_{\text{ISGMR}} &=& 80 \, A^{-1/3} \nonumber \\
E_{\text{IVGDR}} &=& 31.2 \, A^{-1/3} +20.6A^{-1/6}  \\
E_{\text{ISGQR}} &=& 64.7 \, A^{-1/3}. \nonumber
\end{eqnarray}
The values obtained using Eqs.~(\ref{E_GR_syst}) can be compared to predictions provided by the QRPA. The average excitation energy of the GR is obtained by the ratio $M_{1}/M_{0}$, where the $k$-moments are given by
\begin{equation}
\label{M_k}
M_{k}(\hat{Q}_{\lambda \mu}) = \sum _{n} E^{k}_{n} \left | \langle \tilde{O}_{(J^{\pi}=0^{+})} |\hat{Q}_{\lambda \mu}|JM(K)_{n} \rangle \right |^{2}.
\end{equation}
The sum over the QRPA states can be restricted within a given energy interval [$E_{min}$,$E_{max}$] in order to enable comparisons with experiments, which report similarly energy-integrated values.

Table~\ref{tab:90Zr_GR} presents the mean energy values for the isocalar giant monopole/quadrupole resonances (ISGMR/ISGQR) and the isovector giant dipole resonance (IVGDR) for $^{90}$Zr. We compare our QRPA results to the empirical formulae, Eqs.~(\ref{E_GR_syst}), and to experimental values. The energy range used in the QRPA calculation, Eq.~(\ref{M_k}), is chosen to be the same as the one used in the respective experiment. The QRPA calculations exhibit very good agreement for two of the three multipolarities. The IVGDR is found to lie at a little higher energy, $\sim 2$ MeV, in the QRPA calculations when compared to both systematic and experimental values. A similar shift has also been seen in D1M ~\cite{Martini-2016}.
\begin{table}[ht]
\caption{\label{tab:90Zr_GR}Theoretical (QRPA), systematic (Eq.~\ref{E_GR_syst}), and experimental mean energy values for the isocalar giant monopole (ISGMR) and giant quadrupole (ISGQR) resonances and for the isovector giant dipole resonance (IVGDR) for $^{90}$Zr. $E_{\rm{range}}$ represents the energy interval [$E_{min}$,$E_{max}$] summed over. Energies are given in units of MeV.
}
\begin{center}
\begin{tabular}{ c  c  c c c} 
\hline
  GR & QRPA & Syst &  $E_{\rm{range}}$ & Experimental \\
  \hline
  \hline
 ISGMR &  18.6 & 17.9 &[9,36] & 19.17$^{+0.21}_{-0.20}$ \cite{Gupta-2018}  \\
  &  &  &[9,36]  & 17.88$^{+0.13}_{-0.11}$ \cite{Krishichayan-2015} \\
  &  &  &[10,30] & 18.13$^{+0.09}_{-0.09}$ \cite{Gupta-2016} \\
 ISGQR &  15.5   & 14.1 & [9,36] & 14.64$^{+0.22}_{-0.21}$ \cite{Gupta-2018}\\
  &     &  & [9,36] &  14.09$^{+0.20}_{-0.20}$ \cite{Krishichayan-2015}\\
 IVGDR &  19.2  & 16.7 & [0,50] & 16.83$^{+0.04}_{-0.04}$ \cite{Berman-1967}\\
 \hline
\end{tabular}
\end{center}
\end{table}
A comparison of similar quantities, limited to the systematic values, was also carried out for $^{96,98}$Zr.  Results are given in Table~\ref{tab:9698Zr_GR}. We find good agreement between the QRPA results and values from the empirical formulae, with little dependence on deformation. 
\begin{table}[ht]
\caption{\label{tab:9698Zr_GR}Theoretical (QRPA) and empirical (\ref{E_GR_syst}) mean energy values of the ISGMR, ISGQR and IVGDR for $^{96,98}$Zr. $E_{\rm{range}}$ represents the energy interval [$E_{min}$,$E_{max}$] considered. All energies are given in units of MeV.}
\begin{center}
\begin{tabular}{c c  c  c c} 
\hline
  Isotope & GR & QRPA & Syst &  $E_{\rm{range}}$\\
  \hline
  \hline
 $^{96}$Zr& ISGMR &  17.9 & 17.5  &[9,36] \\
 $^{96}$Zr& ISGQR &  15.2   & 13.8  &[9,36] \\
 $^{96}$Zr & IVGDR &  18.4 &  16.4& [0,50] \\
 \hline
 $^{98}$Zr& ISGMR & 17.3  & 17.4 & [9,36] \\
 $^{98}$Zr& ISGQR & 14.95    & 13.7  &[9,36] \\
 $^{98}$Zr & IVGDR &  18.3  & 16.4 & [0,50] \\
 \hline
\end{tabular}
\end{center}
\end{table}
A good measure of the collectivity of individual QRPA states is obtained by calculating their electromagnetic responses.
For low-lying states, it is instructive to investigate reduced transition probabilities, such as the reduced electric quadrupole transition probabilities B(E2;$0^{+}\rightarrow 2^{+}_{1}$).  For the spherical $^{90}$Zr and $^{96}$Zr isotopes, we can compare the QRPA predictions to experimental data, see Table~\ref{BE2}.  We find our results to be in reasonable agreement with the measurements. The values given in Weisskopf units (W.u.) provide information on the relative degree of collectivity of the state. 

We have examined evaluated experimental results for $^{90}$Zr and $^{96}$Zr to obtain insights into deformation and shape coexistence effects.
The energy ratio $E_{4^{+}_{1}}/E_{2^{+}_{1}}$ is expected to be around $2$ for a vibrational spectrum and $\approx 10/3 $ for the band associated with a simple rotor~\cite{Pritychenko-2022}. The experimentally determined energy ratio $E_{4^{+}_{1}}/E_{2^{+}_{1}} = 1.571$ for $^{96}$Zr is smaller than expected for a pure vibration and also far away from the rotational ratio. The systematic study by Pritychenko et al.~\cite{Pritychenko-2022} revealed peculiar aspects of $^{96}$Zr: the $E_{2^{+}_{1}}$ energy is the second highest among all Zr isotopes, and its B(E2;$0^{+}\rightarrow 2^{+}_{1}$) and quadrupole deformation parameter are the smallest among them all despite $^{96}$Zr not being a magic nucleus.
Our QRPA calculations reproduce the fact that the first $2^{+}$ state lies lower in energy in $^{96}$Zr than in $^{90}$Zr. The ratios $E_{4^{+}_{1}}/E_{2^{+}_{1}}$ are also well reproduced by the QRPA calculations. At the same time, our QRPA calculations predict the B(E2;$0^{+}\rightarrow 2^{+}_{1}$) value to be larger for $^{96}$Zr than for $^{90}$Zr, in contrast to the experimental data. Even though the QRPA values are similar for both isotopes, the larger response for $^{96}$Zr indicates a slightly more collective vibrational state. 

\begin{table}[ht]
\caption{Calculated and measured electric B(E2;$0^{+}\rightarrow 2^{+}_{1}$) values.  QRPA results are from this work and adopted values are from Ref.~\cite{Raman-2001,Pritychenko-2016}. The excitation energy is the experimental value. 
\label{BE2}}
\begin{center}
\begin{tabular}{ c |c | c c |c c} 
\hline
  Isotope & $E(2^{+})$  & QRPA              & Exp                  & QRPA & Exp  \\
               & (MeV)       & ($e^{2}b^{2}$) & ($e^{2}b^{2}$)  & (W.u)  & (W.u) \\
  \hline
  \hline
  $^{90}$Zr & 2.186 &0.0381 &0.0610 & 3.2 & 5.1\\
  $^{96}$Zr & 1.750 &0.0470 & 0.055, 0.0314 & 3.6 & 4.2, 2.41\\
  \hline
\end{tabular}
\end{center}
\end{table}

\section{Results}
\label{Results}
\subsection{Electromagnetic responses for selected Zr isotopes}
We can examine the collectivity of excited states further by computing their electromagnetic responses and expressing them as fractions of the relevant energy weighted sum rules (EWSR). The EWSR is obtained by setting $k=1$ in the Eq.~(\ref{M_k}).  We performed the analysis for monopole, dipole, quadrupole, and octupole excitations, considering QRPA excitation energies up to 50 MeV.  In what follows, we discuss results for the $^{90,96,98}$Zr isotopes.

\begin{figure}[htb]
\centering
\includegraphics[scale=0.59]{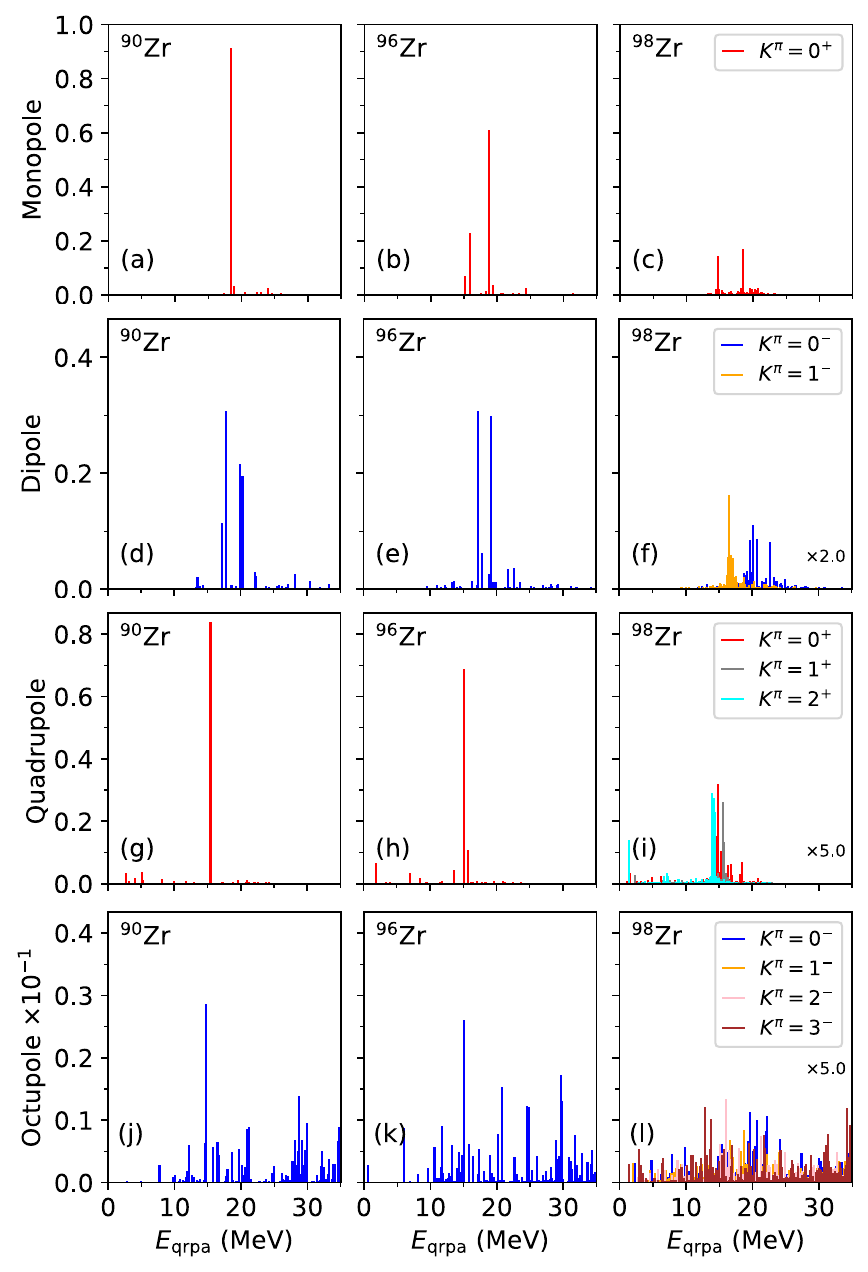}\hfil
\caption{\label{Fraction_EWSR_all} QRPA responses for isoscalar monopole (a-c), isovector dipole (d-f), isocalar quadrupole (g-i) and isovector octupole (j-l) excitations in $^{90}$Zr, $^{96}$Zr, and $^{98}$Zr.  All results are expressed as a fraction of the relevant EWSR, calculated here for energies up to 50 MeV. Results in panels (f), (i) and (l) have been rescaled by a constant value for a better visualization.}
\end{figure}

Figures~\ref{Fraction_EWSR_all}~(a-c) show the isoscalar monopole responses for the $K^{\pi} =0 ^{+}$ QRPA states in the three nuclei. The spherical isotopes (a-b) exhibit a peak at around 19 MeV, which dominates the response  of the EWSR  and exhausts about 60\% for $^{96}$Zr and up to 90\% for $^{90}$Zr.  The strength in the deformed $^{98}$Zr is distributed into two energy regions: the low-energy region near 15 MeV exhausts about half of the EWSR, while the high-energy region at 19 MeV accounts for the remaining strength.

The splitting of the monopole resonance has been predicted and observed in deformed nuclei, with coupling between the monopole resonance and the $K=0$ component of the quadrupole resonance. We will investigate this point in more detail in the next section.  

The responses to the action of the dipole operator are shown in Figs.~\ref{Fraction_EWSR_all}(d-f). The $^{90}$Zr and $^{96}$Zr nuclei exhibit distributions that are similar to each other, with multiple strong peaks occurring around 16-20 MeV where more than 80\% of the EWSR is located. Most of the total dipole strength lies below 20 MeV. 
For the deformed $^{98}$Zr nucleus, contributions from $|K|=0,1$ QRPA excitations are calculated. We observe that the resonance shows a split in energy, which is a well-known characteristic associated with vibrations along the two coordinate axes (long and short) of the spheroidal ground state~\cite{Danos-1958,Okamoto-1958}. The centroid of the $|K|=1$ contribution lies lower in energy than that for the $K =0$ contribution, which is related to the intrinsic oblate deformation of the ground state~\cite{Martini-2016}. 
The isoscalar quadrupole responses are given in Fig.~\ref{Fraction_EWSR_all}(g-i). We observe strong and narrow distributions for the spherical $^{90,96}$Zr nuclei, with about 90\% of the EWSR being concentrated in a single state for each of these cases. A small energy split is observed for the deformed nucleus $^{98}$Zr, with the $K^{\pi}=2^{+}$ peak lying only about 1 MeV below the $K^{\pi}=0^{+}$ peak. This weak quadrupole splitting is in line with what has been observed for other deformed nuclei~\cite{Garg-2018}. For this nucleus, states below 15 MeV exhaust roughly 80\% of the EWSR and 95\% of the EWSR lies below 19 MeV.  Overall, the ISGQR mean energy value is about 15 MeV for the three nuclei studied here, showing narrow distributions and little dependence of centroid energy on the deformation.

Figures ~\ref{Fraction_EWSR_all}~(j-l) show the octupole responses.  Here the situation is very different.  All three nuclei exhibit very fragmented distributions for the octupole operator, with states contributing across a broad range of energies. The contributions from the different K values are difficult to identify, as they are embedded in a dense set of states that contribute.  The effects of deformation on the splitting and fragmentation of the multipole responses discussed here are clearly visible in Fig.~\ref{Fraction_EWSR_all}. For deformed nuclei, we observe that states belonging to $K^{\pi} = 0^{+}$ contribute strongly to both $J = 0$, $J =2$ multipoles, as one can see in panels (c) and (i). Similarly, states belonging to $K^{\pi} = 0^{-}$ and $K^{\pi} = 1^{-}$ contribute to both $J = 1$, $J =3$ multipoles, see panels (f) and (l), even though in this case, the trends are more difficult to distinguish, as the octupole strength is very fragmented. While the trends observed are suggestive of a monopole-quadrupole and a dipole-octupole coupling that occurs through the $K$ components that are present in the two paired multipoles, a more detailed analysis is desirable. We will focus on this aspect in the next section.

\begin{figure}[ht]
\centering
\includegraphics[scale=0.6]{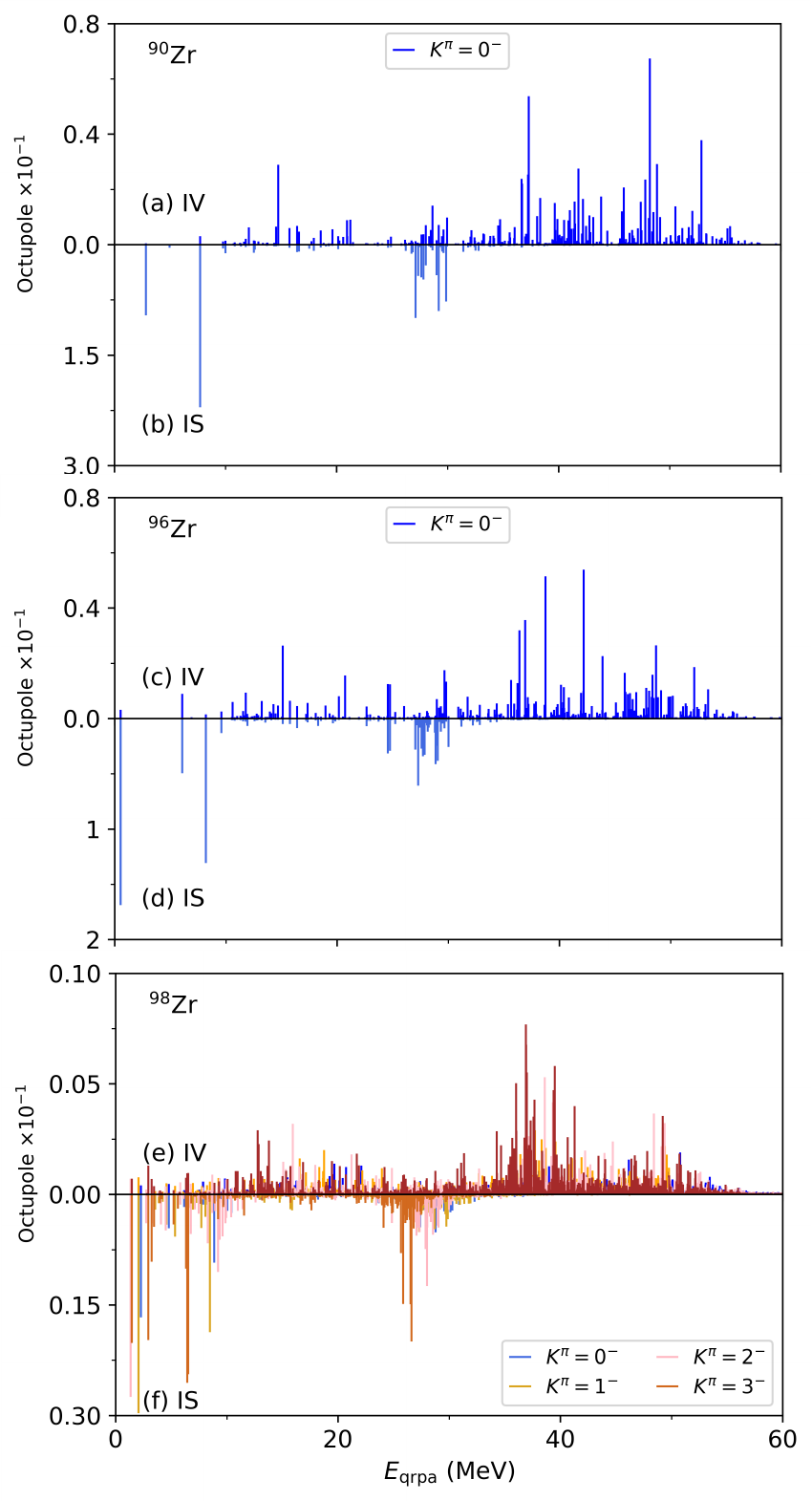}
\caption{\label{Fraction_EWSR_J_3_only}Octupole response as a fraction of EWSR for $^{90}$Zr (top), $^{96}$Zr (middle) and $^{98}$Zr (bottom). Both Isovector (IV) and Isoscalar (IS) components are shown. The IV response is always shown above the 0 horizontal reference line, while the IS response is plotted bellow it.}
\end{figure}

The octupole response is clearly more complex than the responses associated with the lower multipolarities. Octupole responses are expected to be split into a low-energy isoscalar (IS) and high-energy isovector (IV) component \cite{Harakeh-2001,Goutte-2008}. To investigate the octupole excitations further, we show both the isoscalar and isovector responses in Fig.~\ref{Fraction_EWSR_J_3_only}, for excitation energies up to 60 MeV.  For each isotope, the isoscalar response is given in the lower part and the isovector response is shown in the upper part. Above about 10 MeV excitation energy, the patterns for all three isotopes are similar.  There are strong isoscalar contributions around 25-30 MeV and the isovector response dominates in the energy regime of 30-50 MeV. Below 10 MeV, we observe strong isoscalar contributions located in a few, well-separated, low-energy peaks for the spherical nuclei, and a more spread-out distribution in the deformed $^{98}$Zr nucleus. The visual analysis of the deformed case becomes difficult due to the large number of contributions from states with $|K|^{\pi} = 0^{-}, 1^{-},2^{-}$ and $3^{-}$. 

\subsubsection{Transition densities: monopole-quadrupole and dipole-octupole coupling in $^{98}$Zr}
\label{QRPA_results_coupling}

The isoscalar giant monopole resonance has been extensively studied, in particular for spherical nuclei.  It is viewed as a breathing mode that can provide valuable information on the incompressibility of nuclear matter.  Its properties in deformed nuclei, however, are less well understood.  It is known (see e.g. \cite{Martini-2016, Goutte-2008, Peru-2014}) that appreciable mixing can occur between the ISGMR and the $K$=0 component of the ISGQR, which leads to a redistribution of the monopole strength - an effect we have observed in our results shown in Figure~\ref{Fraction_EWSR_all}.
To shed some light on the issue, we analyze the structure of selected $K^{\pi}$=$0^{+}$ QRPA states that contain both monopole and quadrupole strength.

\begin{figure}[htb]
\centering
\includegraphics[scale=0.58]{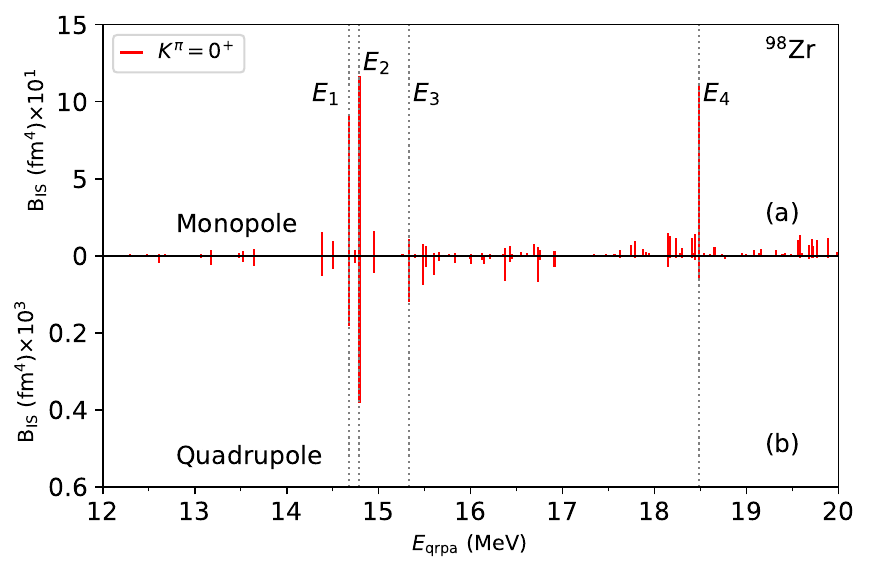}
\caption{\label{98Zr_Mon_Qua}Isoscalar monopole (a) and quadrupole (b) responses for $^{98}$Zr for $K^{\pi}=0^{+}$ QRPA states. The vertical dotted lines indicate states with strong responses for both of electromagnetic modes. Their radial transition densities are depicted in Fig.\ref{98Zr_dens_J_02_K_0}.}
\end{figure}

\begin{figure}[htb]
\centering
\includegraphics[scale=0.56]{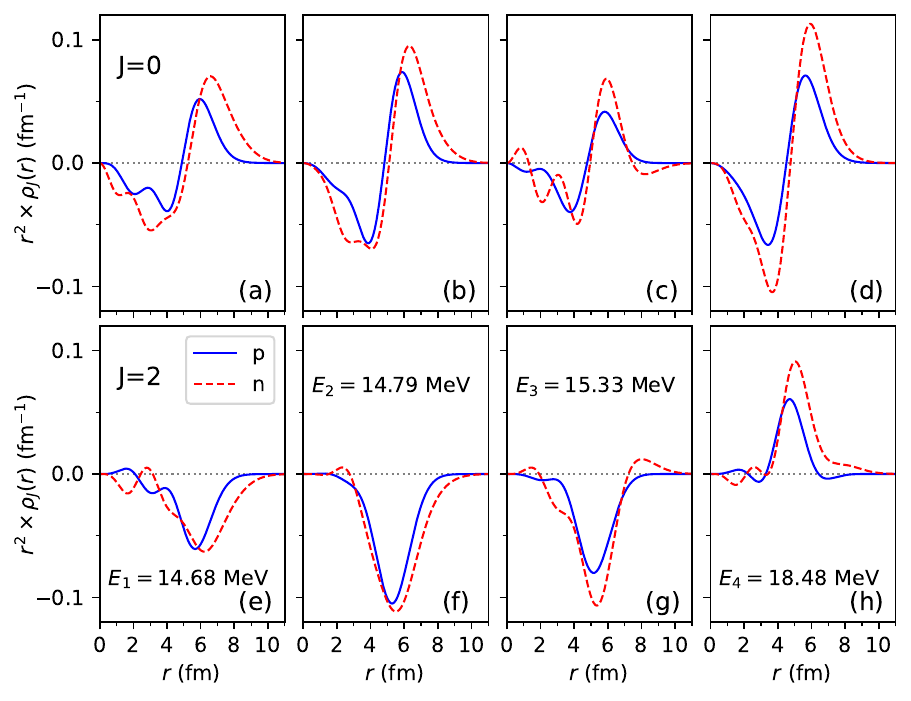}
\caption{\label{98Zr_dens_J_02_K_0} Radial projection ($J=0$ upper panels and $J=2$ lower panels) of $^{98}$Zr transition densities for the four $K^{\pi}=0^{+}$ QRPA states selected from Fig. \ref{98Zr_Mon_Qua}.}
\end{figure}

Figure~\ref{98Zr_Mon_Qua} shows the $K^{\pi}=0^{+}$ monopole and quadruple responses in the energy regime $E_{\rm{qrpa}}=$12-20 MeV, i.e. around the ISGMR and ISGQR peaks.  We select four states that strongly contribute to both the monopole (upper part) and quadrupole (lower part) responses, at $E_1$ = 14.68 MeV, $E_2$ = 14.79 MeV, $E_3$ = 15.33 MeV, and $E_4$ = 18.48 MeV.  These are indicated in the figure by dotted vertical lines.
Radial transition densities (defined in Eq.~\ref{rho_LSJ}) associated with these four states, for both $J=0$ and $J=2$, are shown in Fig.~\ref{98Zr_dens_J_02_K_0}.
Proton and neutron transition densities are indicated by solid blue and dashed red curves, respectively.
The predominantly isoscalar character of these excitations is clearly visible - proton and neutron densities are in phase at almost all radial points.
Panels (a-d) show patterns of positive (increased) density around $6$ fm and negative (reduced) density near $3$ fm, i.e. all four states are characterized by a breathing-mode type of excitation in the monopole sector. 
The radial transition densities for the quadrupole sector, in contrast, are concentrated on the surface of the nucleus, with little or no compensating density change at smaller radii, see panels (e-h). 
The transition densities for both modes have peaks near the nuclear surface. We can therefore expect that these modes will be excited in direct nuclear reaction experiments, such as inelastic (hadronic) scattering.

\begin{figure}[htb]
\centering
\includegraphics[scale=0.58]{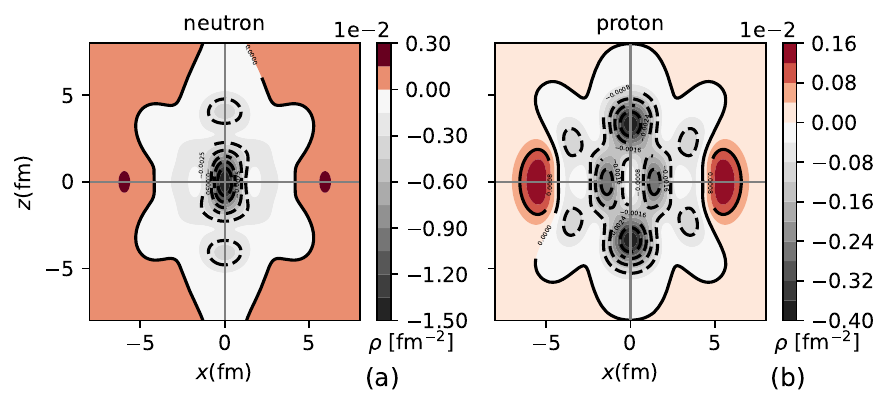}
\includegraphics[scale=0.58]{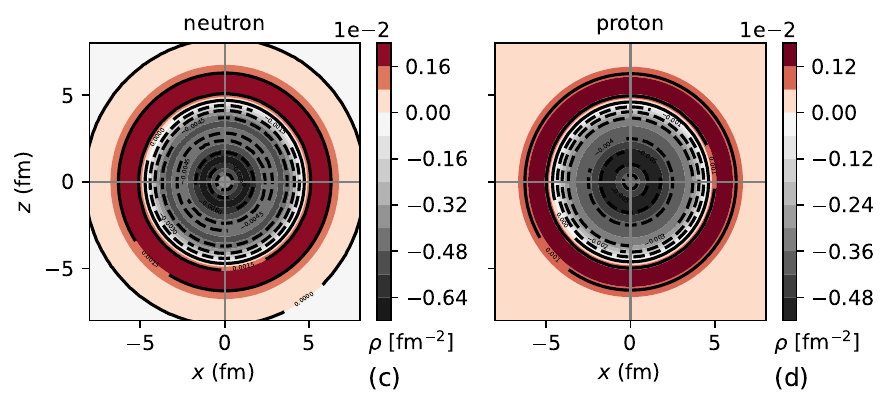}
\includegraphics[scale=0.58]{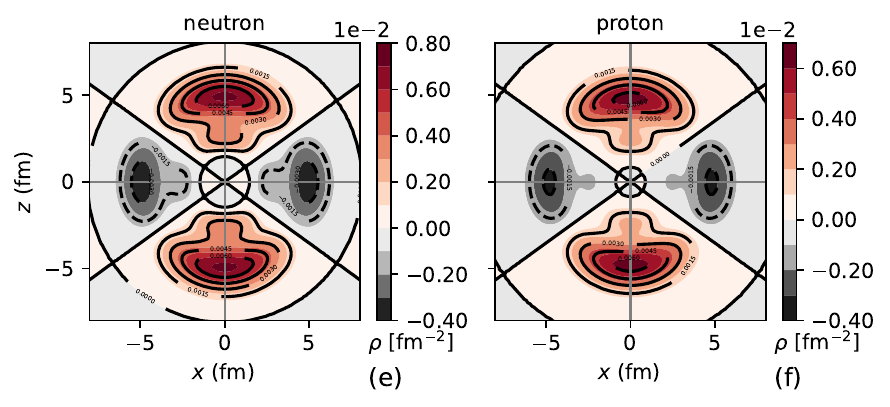}
\caption{\label{98Zr_qrpa_int_td} Intrinsic neutron and proton transition densities for selected $K^{\pi} = 0^{+}$ states in $^{98}$Zr and $^{96}$Zr. Panels (a) and (b) show neutron and proton transition densities for the excited $K^{\pi}=0^{+}$ state at $E_{1}= 14.68$ MeV in $^{98}$Zr. 
For comparison, intrinsic transition densities for a pure  $K^{\pi}=0^{+}, J=0$ state at $E_{\rm{qrpa}}= 18.74$ in the spherical nucleus $^{96}$Zr are shown in panels (c) and (d).  This state is located near the peak of the GMR in that nucleus.
Similarly, panels (e) and (f) show intrinsic transition densities for a pure  $K^{\pi}=0^{+}, J=2$ state at $E_{\rm{qrpa}}= 15.08$ in $^{96}$Zr.  This state is located near the peak of the GQR in $^{96}$Zr. The transition densities for the deformed nucleus $^{98}$Zr exhibit rich excitation patterns which combine contributions from monopole, quadrupole, and higher multipole excitations.}
\end{figure}

Each of the states discussed contains both $J=0$ and $J=2$ components, plus possibly higher multipolarities.  A more complete, and also more difficult to interpret, picture is given by the intrinsic transition densities (defined in Eq.~\ref{td}), which contain all these components.  
In Fig.~\ref{98Zr_qrpa_int_td}, we show the intrinsic transition density for the $E_{1}= 14.68$ MeV state.  The neutron density is given in panel (a) and the proton density is in panel (b).  For comparison, we have also included intrinsic transition densities for the strongest monopole and quadrupole excitations in the neighboring spherical nucleus $^{96}$Zr in the next two rows of the figure.  Pure monopole excitations are shown in panels (c-d), and quadrupole excitations are presented in panels (e-f).
The transitions in the spherical nucleus demonstrate the breathing and surface oscillation patterns associated with monopole and quadrupole resonances, respectively. Both neutrons and protons transition densities in panel (a) and (b), respectively, show monopole and quadrupole components.  Some characteristics of the breathing mode (opposite signs for the densities at the nuclear surface and the interior) are still visible. There are also strong oscillations discernible at the surface of the nucleus; these exhibit quadrupole character.  The shapes look quite complex, as they contain contributions from spherical harmonics of multiple orders.

\begin{figure}[htb]
\centering
\includegraphics[scale=0.58]{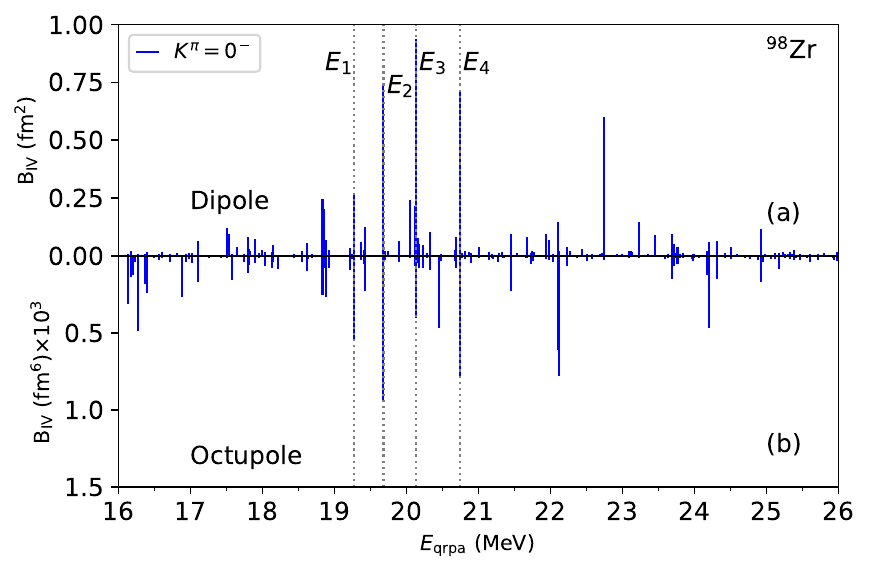}
\caption{\label{98Zr_Dip_Oct}Isovector dipole (a) and octupole (b) responses of $K=0^{-}$ QRPA states for $^{98}$Zr. The vertical dotted lines represent states with strong response for both electromagnetic modes.}
\end{figure}

It is known that the isovector giant dipole resonance (IVGDR) is split into two components in axially-symmetric deformed nuclei and that the octupole response has multiple components, but details of the coupling between these modes are not well studied.  Here we start by investigating the coupling of these modes through the $K^{\pi} = 0^{-}$ component.  We focus on the isovector responses in the energy regime $E_{\rm{qrpa}}= 16-26$ MeV, see Fig.~\ref{98Zr_Dip_Oct}. 
In analogy to the previous case, we identify four $K^{\pi} = 0^{-}$ QRPA states with strong B(E1) and B(E3) values, at $E_1$ = 19.27 MeV, $E_2$ = 19.68 MeV, $E_3$ = 20.13 MeV, and $E_4$ = 20.74 MeV.  Vertical lines mark these four states in the figure.

\begin{figure}[htb]
\centering
\includegraphics[scale=0.56]{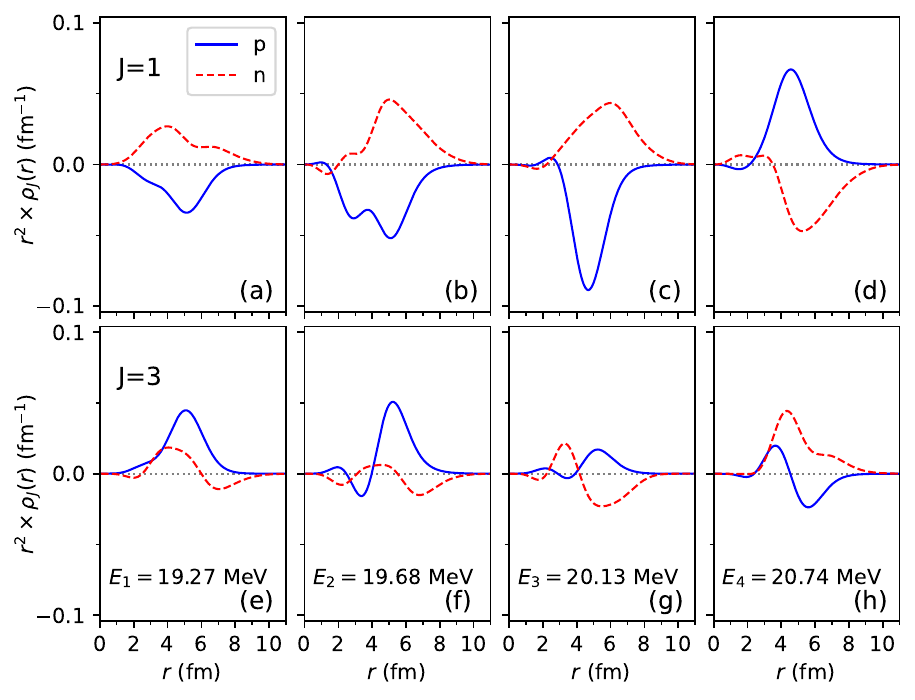}
\caption{\label{98Zr_dens_J_13_K_0} Radial projection ($J=1$ upper panels and $J=3$ lower panels)  of the transition density for the $K=0^{-}$ QRPA states selected in Fig. \ref{98Zr_Dip_Oct}.}
\end{figure}

Radial transition densities associated with these four states, for both $J=1$ and $J=3$, are shown in the upper and lower panels of Fig.~\ref{98Zr_dens_J_13_K_0}, respectively.
Proton and neutron transition densities, indicated by solid blue and dashed red curves, respectively, oscillate out of phase in the upper panels, demonstrating the isovector nature of the dipole resonance.
A similar isovector behavior is seen for the octupole modes in the lower panels, but the shapes of the radial densities are more complicated than those for the dipole mode.
As in the monopole-quadrupole case discussed above, the transition densities exhibit peaks near the nuclear surface, i.e. these modes can be excited in direct inelastic scattering with hadrons.


\begin{figure}[htb]
\centering
\includegraphics[scale=0.58]{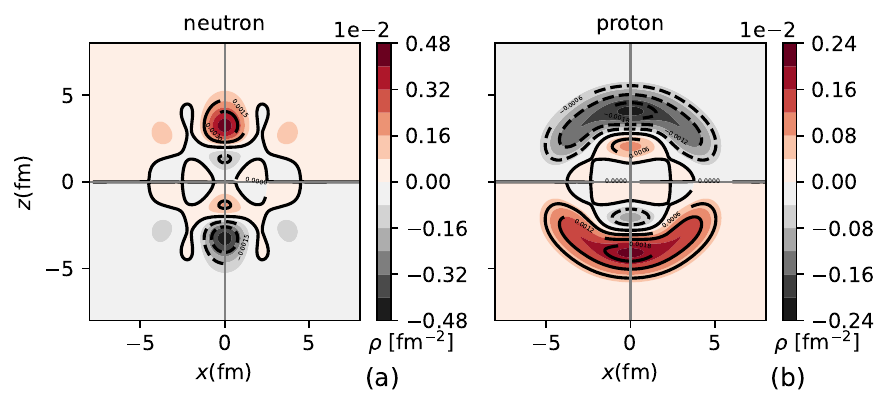}
\includegraphics[scale=0.58]{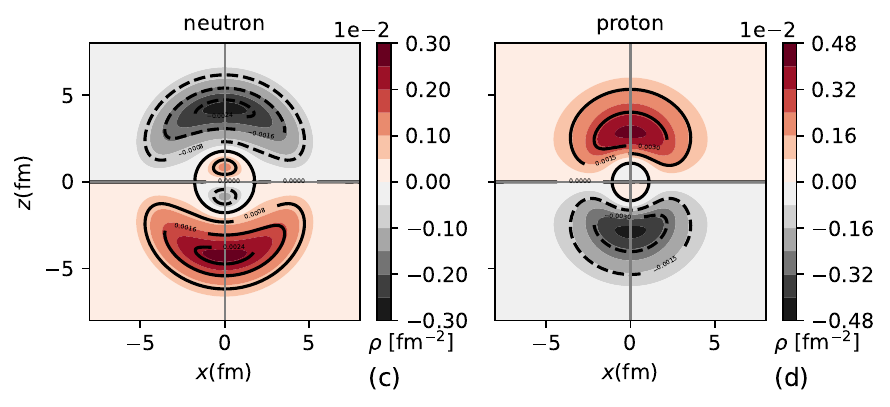}
\includegraphics[scale=0.58]{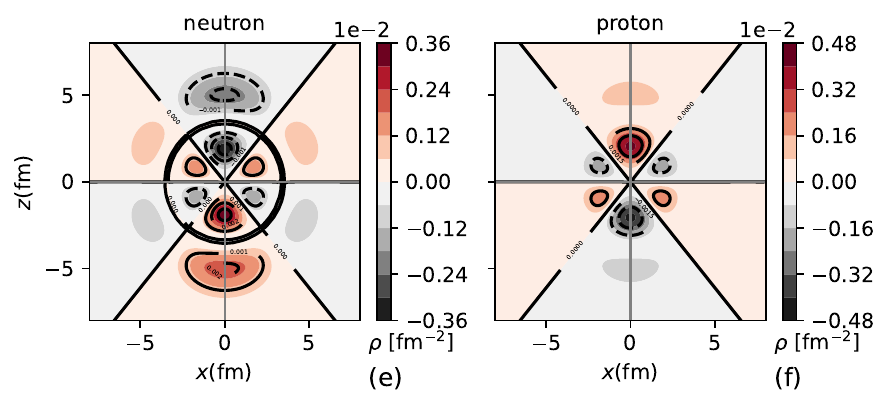}
\caption{\label{98Zr_qrpa_int_0m_td_316} 
Intrinsic neutron and proton transition densities for selected $K^{\pi} = 0^{-}$ states in $^{98}$Zr and $^{96}$Zr.
Panels (a) and (b) demonstrate the coupling between isovector dipole and octupole states in the deformed $^{98}$Zr nucleus.
Shown are neutron and proton transition densities for the excited $K^{\pi} = 0^{-}$ state at $E_{3}= 20.13$ MeV, which has contributions from dipole, octupole, and higher multipole excitations.
For comparison, intrinsic transition densities for a pure  $K^{\pi}=0^{-}, J=1$ state at $E_{\rm{qrpa}}= 17.22$ MeV in the spherical nucleus $^{96}$Zr are shown in panels (c) and (d).  This state represents the strongest isovector dipole peak in Fig.~\ref{Fraction_EWSR_all} (e)
Similarly, panels (e) and (f) show intrinsic transition densities for a pure  $K^{\pi}=0^{-}, J=3$ state at $E_{\rm{qrpa}}= 15.12$ MeV in $^{96}$Zr.  This state represents a very strong isovector octupole peak in Fig.~\ref{Fraction_EWSR_all} (k).
}
\end{figure}

In panels (a) and (b) of Fig.~\ref{98Zr_qrpa_int_0m_td_316}, we show the intrinsic transition density for the $K^{\pi} = 0^{-}$ state at $E_{3}= 20.13$ MeV.  The proton sector exhibits a strong dipole shape, while the neutron transition density is more complicated, due to the impact of the octupole (and possibly higher multipole) contributions: a shaded cloud is observable near the surface of the nucleus. The intrinsic transition shapes presented in the upper panels can be compared to the pure dipole and octupole transition densities associated with states in the neighboring spherical $^{96}$Zr nucleus.  These are shown in the next two rows of the figure.
The isovector dipole excitations, shown in panels (c-d), have an easily recognizable pattern that is associated with the $Y_{10}$ spherical harmonic.  The dominant features of this pattern is similar for both protons and neutrons. The isovector octupole excitations, shown in panels (e-f), have a more complicated pattern, since they are following the shape of the $Y_{30}$ function.  The four-portion oscillations ($J=2$) in the inserts (e) and (f) of Fig.~\ref{98Zr_qrpa_int_td} are replaced by six ($J=3$) in Figure \ref{98Zr_qrpa_int_0m_td_316}, with the number of portions increasing with multipolarity. Similarly to only one portion for monopole and two for dipole.

\begin{figure}[htbp]
\centering
\hspace{-0.46cm}\includegraphics[scale=0.61]{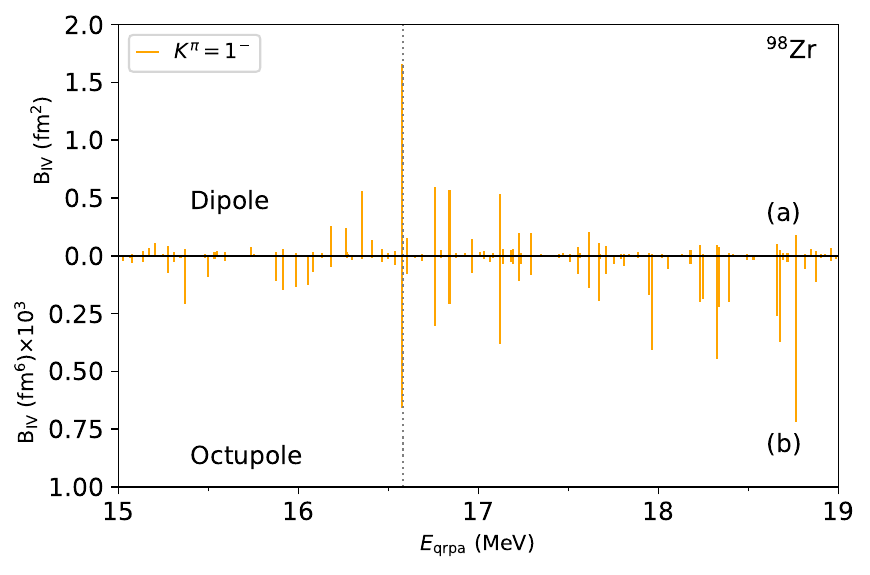}
\caption{\label{98Zr_Dip_Oct_K=1m_zoom}Isovector dipole (a) and octupole (b) responses of $K^{\pi}=1^{-}$ QRPA states for $^{98}$Zr. The vertical dotted line at $E_1$ =  16.58 MeV, indicates a state with strong response for both electromagnetic modes.}
\end{figure}

\begin{figure}[htbp]
\centering
\includegraphics[scale=0.84]{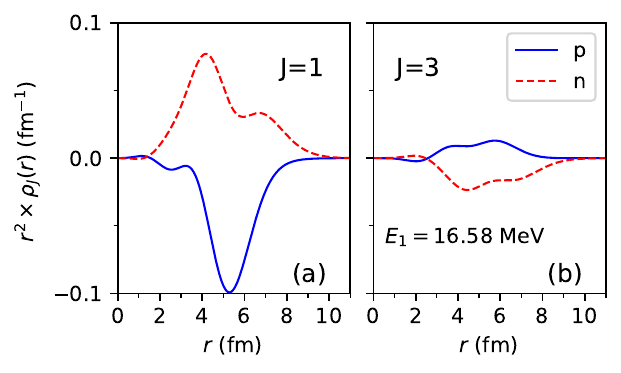}
\caption{\label{98Zr_dens_J_13_K_1} Radial projection ($J=1$ left and $J=3$ right) of the transition density for the $K^{\pi}=1^{-}$ QRPA state at $E_1$ =  16.58 MeV in $^{98}$Zr.}
\end{figure}

\begin{figure}[htbp]
\centering
\includegraphics[scale=0.57]{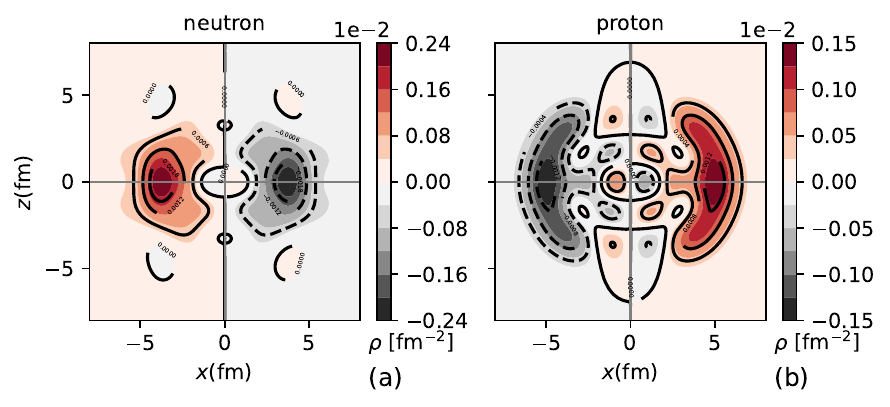}
\caption{\label{98zr_qrpa_int_1m_td_201} Intrinsic transition densities for the selected QRPA state with $K=1^{-}$ at $E_{\rm{qrpa}}= 16.58$ MeV in $^{98}$Zr (see Fig.\ref{98Zr_Dip_Oct_K=1m_zoom}).}
\end{figure}

Dipole and octupole responses can also couple through the $K^{\pi} = 1^{-}$ QRPA components in deformed nuclei.  Here, we study this coupling for a selected state in the deformed $^{98}$Zr nucleus. We focus on the isovector responses in the energy regime $E_{\rm{qrpa}}= 15-19$ MeV. Fig.~\ref{98Zr_Dip_Oct_K=1m_zoom} shows both the B(E1) and B(E3) computed for the $K^{\pi} = 1^{-}$ component. A vertical, dotted line marks the $K^{\pi} = 1^{-}$ state at $E_1$ = 16.58 MeV, which has strong dipole and octupole contributions.
Radial transition densities associated with this state, for both $J=1$ and $J=3$, are shown in Fig.~\ref{98Zr_dens_J_13_K_1}.
Proton and neutron transition densities, indicated by solid blue and dashed red curves, respectively, oscillate out of phase in both panels, demonstrating the isovector nature of the two excitations.
The $J=1$ amplitudes are significantly larger than those for $J=3$. The dominance of the $J=1$ oscillations is reflected in intrinsic transition densities for this state that maintain their characteristic dipole form, as seen in Fig.~\ref{98zr_qrpa_int_1m_td_201}.  The proton transition density clearly resembles a dipole-like response, but with a 90 degree rotation relative to the cases discussed above.  This is due to the fact that the density was calculated from the $K^{\pi} = 1^{-}$ component. 

The implications of the coupling between collective excited modes for experimental observations need to be considered. Monopole-quadrupole coupling affects the determination of nuclear incompressibility, as previously recognized ~\cite{Bahini-2022}. We demonstrated a similar coupling between the isovector dipole and octupole excitations through the $K^{\pi}=0^{-}$ and $K^{\pi}=1^{-}$ QRPA components. The impact of this coupling on isovector dipole excitations studied experimentally also requires further investigation, as the electromagnetic dipole resonance plays an important role in calculating neutron capture cross-sections for nuclear astrophysics and other applications.



\section{\label{summary} Summary and Outlook }
The interplay between different multipolarities in the QRPA modes has been studied for the deformed $^{98}$Zr isotope. We have benchmarked our consistent HFB+QRPA approach for $^{90}$Zr and used the $^{96-98}$Zr isotopes to explored how the spherical-deformed shape transition impacts properties of excited states. The QRPA low-lying states, as we well as giant resonances predicted here are found to compare favorably with available experimental data and known systematics. Our exact treatment of the Coulomb interaction, in contrast with approximations used in the fitting of the D1S and D1M parameterization of the Gogny interaction, has led us to observe an energy shift in the predicted excitation energy for the first $3{-}$ state of $^{96}$Zr. We found that the complete treatment can reduce the pairing energies of the protons in agreement with previous results obtained for heavier systems \cite{Rod2022, Rod2023}. Near shell closures, the gap between occupied and unoccupied single-particle levels is expected to be large enough to significantly suppress pairing correlations \cite{Peru-2000,Anguiano-NPA-2001}. Large pairing energy values reduce the nuclear binding energy difference between open-shell and closed-shell configurations by softening the potential energy curves.
 
For the  deformed $^{98}$Zr nucleus, the coupling between the isoscalar monopole and quadrupole excitations through their $K^{\pi}=0^{+}$ component was shown for representative QRPA states. The intrinsic transition densities obtained for these cases exhibit excitation configurations with protons and neutrons oscillating in phase and spatial patters that combine contributions from both $J =0$ and $J =2$ multipolarities. A feature confirmed by the large amplitudes of the respective projected radial transition densities. The monopole-quadrupole coupling shown here adds to previous studies in different nuclei and contributes to a better understanding of such phenomena, which may impact the determination of the compressibility and the nuclear equation of state \cite{Garg-2018,Bahini-2022,Blaizot-1980, RocaMaza-2018, Xuwei2022, Adri2023}. Notably, in this work, we discussed in detail the complex scenario of the dipole-octupole coupling present in both $K^{\pi} = 0^{-} $ and $K^{\pi} = 1^{-}$ states of $^{98}$Zr. We have identified QRPA states with strong dipole and octupole contributions in the energy range 16-26 MeV for $K^{\pi} = 0^{-}$ and at $15-19$ MeV for $K^{\pi} = 1^{-}$. Our findings are in agreement  with previous studies by Yoshida and Nakatsukasa who employed a Skyrme-functional approach for Nd and Sm isotopes  \cite{Yoshida-2013} to show that axial deformation enables mixing and impacts the width and strength of dipole responses. The QRPA transition densities presented here further elucidate the microscopic picture of this coupling in deformed nuclear systems in a framework that uses the finite-range Gogny interaction. Their intrinsic transition densities exhibit a clear resemblance with dipole- and octuple-like spatial shapes. Protons and neutrons move mostly in opposite phase, resulting on a large isovector contribution from $J =1$ and $J=3$ multipolarities. The observed coupling of dipole and octupole strength in our QRPA results arises, similarly to the monopole-quadrupole case, from the deformation-induced mixing of different multipolarity values within the QRPA state. The multicomponent coupling in $K^{\pi} = 0^{-}, 1^{-}$  has not, to the best of our knowledge, been fully explored before. We recall that accurate reaction cross section predictions rely heavily on gamma-ray EJ-strength functions derived from experimental data and on transition densities obtained from nuclear structure models. We believe our findings stress the importance of such coupling in the deformed nuclei and will lead to further studies of their impact on observed quantities.

Overall, the present study has produced new insights regarding the coupling between collective modes in deformed $^{98}$Zr utilizing a consistent HFB-QRPA and Gogny D1M force approach and motivates additional studies. Investigations of couplings across a broader set of isotopes, from different parts of the nuclear chart, would be interesting. Given the recent results by Porro et al. \cite{Porro-2024} regarding the exact projection after variation with appropriate treatment of spurious rotational states, monopolar quadrupole coupling still appears to occur. This statement must be confirmed by an exact projection before  variation which is not yet implemented. Ongoing work aims at integrating the structure information contained in the QRPA transition densities into direct-reaction calculations. This will directly connect the microscopic structure predictions with experimental observables, in particular for charged-particle inelastic scattering experiments. In addition, our efforts are aiming at the inclusion of rotational degrees of freedom, moments of inertia and associated excitations.


\clearpage 

\appendix
\section{HFB implementation and numerical convergence}
\label{HFB_implementation_and_numerical_convergence}
In our HFB implementation we perform the variation calculation using the matrix elements of the density $\rho$ matrix and pairing $\kappa $ tensor. We refer to Refs.\cite{Ring-2004,Younes-2009,Younes-2019} for further details about matrix elements expressions and a more complete description of both the HFB theory and the mean field algorithm employed here. Our effective interaction of choice consists of the finite-range Gogny force~\cite{Decharge-1980} and the Coulomb interaction. The D1M and D1S parameterizations, which are of interest to the present study can be found, for example in \cite{Goriely-2009, Younes-2019}. Unless otherwise stated, we use the D1M parameters and include the Coulomb contribution with its exact exchange part without approximations and we apply a one-body correction for the center-of-mass problem \cite{Younes-2019}.

\begin{figure}[htb]
\centering
\includegraphics[scale=0.57]{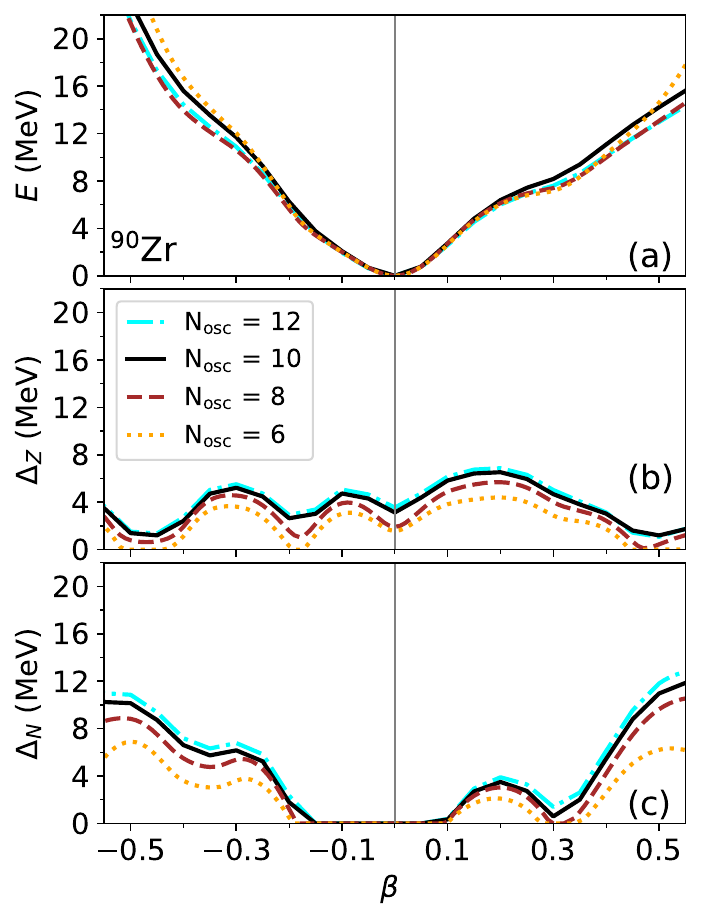}
\caption{\label{90zr_n6n8n10n12}Convergence of HFB solutions for $^{90}$Zr as function of the size of the harmonic oscillator basis. Panel (a) shows the energy as function of deformation parameter, and panels (b) and (c) give the energies of the proton and neutron pairing fields, respectively. Increasing the model space to include major oscillator shells with $N_{\rm{osc}} >10$ results in only minor changes. The energy curves (a) are shifted in energy to align their individual minima at the origin.}
\end{figure}

To demonstrate the convergence of our results, we conducted calculations for $^{90}$Zr using four different basis sizes. Figure~\ref{90zr_n6n8n10n12} shows results for $^{90}$Zr performed with four different basis sizes, $N_{\rm{osc}} = 6, 8, 10, 12$. Panel (a) illustrates the energy profile, with energies shifted for each individual curve to align the minima for a better comparison, while panels (b) and (c) show the pairing energies for protons and neutrons, respectively.
The absolute energy values are given in Table~\ref{Table:90zr_n6n8n10n12_energy}. The HFB solutions are stable in a basis with 10 major oscillator shells, with the inclusion of two additional shells providing only minor changes to the ground state energy.  
\begin{table}[htb]
\caption{\label{Table:90zr_n6n8n10n12_energy} HFB binding energy for $^{90}$Zr as function of the size of harmonic oscillator basis.}
\begin{center}
\begin{tabular}{c | c c c c} 
\hline
$N_{\rm{osc}}$  & 6 & 8 & 10 & 12 \\
\hline
$E_{\rm{GS}}$ (MeV) & -780.50 & -786.78&  -791.60 & -791.85\\
\hline
\end{tabular}
\end{center}
\end{table}
Table \ref{tab:Zr_isotopes_TW} summarizes the ground state properties of Zr isotopes obtained in this work. For comparison, we also include the experimentally evaluated charge radius values.
\begin{table}[htb]
\caption{\label{tab:Zr_isotopes_TW}Ground state properties of Zr isotopes from this work, including the deformation parameter $\beta$, the HFB ground state energy $E_{\rm{GS}}$, the proton and neutron pairing energies, $\Delta_{Z}$ and $\Delta_{N}$, and the root-mean-square nuclear and charge radii $R_{\rm{rms}}$ and $R_{\rm{c}}$, respectively. Experimental charge radii $R^{\rm{exp}}_{\rm{c}}$ are from \cite{Nudat, Angeli-2013}.}
\begin{center}
\begin{tabular}{c c c c c c c c c } 
\hline
Nuc. & $\beta $ & $E_{\rm{GS}}$  & $\Delta_{Z}$ & $\Delta_{N}$ & $R_{\rm{rms}}$ & $R_{\rm{c}}$ & $R^{\rm{exp}}_{\rm{c}}$ \\
             &               &   (MeV)              &  (MeV)   & (MeV) &  (fm)                  &   (fm)       &   (fm)        \\
\hline 
\hline
$^{90}$Zr       & 0.0  & -791.60      & 3.14         & 0.00 &4.20        & 4.22      & 4.2694$\pm$ 0.001            \\
$^{96}$Zr       & 0.0   & -832.41      & 1.30         & 9.31& 4.31         & 4.28   & 4.3512$\pm$ 0.0015                \\
$^{98}$Zr       & -0.2          & -844.47     &1.80          & 9.28 & 4.37        & 4.33   & 4.4012$\pm$ 0.0164               \\
\hline 
\end{tabular}
\end{center}
 \end{table}
 
\section{Angular momentum restoration}
\label{Angular_momentum_restoration}

In this section, we present a short review of the angular momentum restoration employed in our calculations. One interprets this procedure as the transformation of the states, transition amplitudes, and matrix elements, from the intrinsic frame (where calculation are performed) to the laboratory frame (the space where measurements are performed).
QRPA states with good angular momentum $ |JM(K)_{n} \rangle $ are obtained from the intrinsic, deformed, QRPA states $|\theta _{n},K \rangle$ using angular momentum restoration~\cite{Peru-2014}:
\begin{eqnarray}
\label{Psi_n}
  |JMK_{n} \rangle &=& \frac{\sqrt{2J+1}}{4\pi}\int d\Omega \mathcal{D}^{J}_{MK}(\Omega )\hat{R}(\Omega )|\theta _{n},K \rangle . 
\end{eqnarray}
Here $\mathcal{D}^{J}_{MK}$ and $\hat{R} (\Omega)$ are the Wigner $\mathcal{D}$-matrix and the rotation operator, respectively.
The ground state of an even-even nucleus is related to the intrinsic (deformed) ground state, $|0_{\text{def}},(K=0) \rangle \equiv |0_{\text{def}} \rangle$, via
\begin{eqnarray}
\label{Psi_0}
  |\tilde{O}_{(J^{\pi}=0^{+})} \rangle = \frac{1}{2\pi} \int d\Omega \mathcal{D}^{0}_{00}(\Omega )\hat{R}(\Omega) |0_{\text{def}} \rangle .
\end{eqnarray}
Here, we are interested in the response of the system to an external field, such as the electromagnetic operator (the derivations here remain true for any spherical tensor)
\begin{eqnarray}
  \hat{Q}_{\lambda \mu} = r^{\lambda }Y_{\lambda \mu}.
\end{eqnarray}
Operators written in the laboratory frame (LAB) can be expanded as
\begin{eqnarray}
\label{QLAB}
  \hat{Q}^{\text{LAB}}_{\lambda \mu} = \sum _{\mu ^{\prime}}\mathcal{D}^{*\lambda}_{\mu \mu ^{\prime}}(\Omega )\hat{Q}^{\text{INT}}_{\lambda \mu^{\prime}}= r^{\lambda } \sum _{\mu ^{\prime}}\mathcal{D}^{*\lambda}_{\mu \mu ^{\prime}}(\Omega ) Y_{\lambda \mu ^{\prime}}(\hat{r}). \nonumber \\
\end{eqnarray}
where (INT) represents the operator acting in the intrinsic frame.  The transformation of the operator is given by a simple sum over the components of the operator in the intrinsic frame, weighted by the Wigner $\mathcal{D}$-matrix \cite{Edmonds-74}. From now one, we make use of this expansion and omit the label indicating the relevant frames. Combining the three pieces (Eqs.~(\ref{Psi_0}-\ref{QLAB})), we obtain 
\begin{widetext}
\begin{eqnarray*}
  \langle  JMK_{n}|\hat{Q}_{\lambda \mu}|\tilde{O}_{(J^{\pi}=0^{+})} \rangle &=& \frac{\sqrt{2J+1}}{8\pi^{2}} \sum _{\mu ^{\prime}} \int d\Omega _{1}\int d\Omega _{2}  \langle \theta _{n},K| \mathcal{D}^{J*}_{MK}(\Omega _{1} )\hat{R}^{\dagger}(\Omega _{1}) r^{\lambda }\mathcal{D}^{*\lambda}_{\mu \mu ^{\prime}}(\Omega _{2}) Y_{\lambda \mu ^{\prime}}(\hat{r}) \mathcal{D}^{0}_{00}(\Omega _{2})\hat{R}(\Omega _{2})| 0_{\text{def}} \rangle
\end{eqnarray*}
\end{widetext}
This expression can be simplified if we assume that the overlap of the integrals $\int d\Omega _{1}\int d\Omega _{2}$ vanishes unless $\Omega _{1}=\Omega _{2} =\Omega$ holds. This is sometimes referred as needle approximation~\cite{Ring-2004,Peru-2014}. It allows us to trivially perform one angular integral to obtain 
\begin{widetext}
\begin{eqnarray*}
  \langle  JMK_{n}|\hat{Q}_{\lambda \mu}|\tilde{O}_{(J^{\pi}=0^{+})} \rangle &=&\frac{\sqrt{2J+1}}{8\pi ^{2}} \sum _{\mu ^{\prime}}  \int d\Omega  \langle \theta _{n},K | \mathcal{D}^{J*}_{MK}(\Omega )\hat{R}^{\dagger}(\Omega ) r^{\lambda }\mathcal{D}^{*\lambda}_{\mu \mu ^{\prime}}(\Omega ) Y_{\lambda \mu ^{\prime}}(\hat{r}) \mathcal{D}^{0}_{00}(\Omega )\hat{R}(\Omega )| 0_{\text{def}} \rangle
\end{eqnarray*}
\end{widetext}
Since $\hat{R}^{\dagger}(\Omega)\hat{R}(\Omega) =1$, and by evaluating the angular integral using Eqs. (4.2.7) and (4.6.2) in Edmonds~\cite{Edmonds-74}
\begin{widetext}
\begin{eqnarray*}
  \int d\Omega \mathcal{D}^{J*}_{MK}(\Omega )\mathcal{D}^{*\lambda}_{\mu \mu ^{\prime}}(\Omega ) \mathcal{D}^{0}_{00}(\Omega )&=& (-1)^{\mu - \mu^{\prime}}\left ( \int d\Omega \mathcal{D}^{J}_{MK}(\Omega )\mathcal{D}^{\lambda}_{-\mu -\mu ^{\prime}}(\Omega ) \mathcal{D}^{0}_{00}(\Omega ) \right )^{*} \\
  &=&8\pi^{2}\,(-1)^{\mu -\mu ^{\prime}} 
 \begin{pmatrix}
  J  & \lambda & 0\\
  M  &  -\mu  & 0 
  \end{pmatrix}
  \begin{pmatrix}
 J & \lambda &0 \\
 K & -\mu ^{\prime}& 0 
  \end{pmatrix},
\end{eqnarray*}
\end{widetext}
we finally obtain the following expression for any spherical tensor operator $\hat{Q}_{\lambda \mu}$:
\begin{widetext}
\begin{eqnarray*}
\langle JMK_{n}|\hat{Q}_{\lambda \mu}|\tilde{O}_{(J^{\pi}=0^{+})} \rangle &=& \sqrt{2J+1} \sum _{\mu ^{\prime}}(-1)^{\mu^{\prime} -\mu } 
 \begin{pmatrix}
J  & \lambda &0 \\
M  &  -\mu & 0 
\end{pmatrix}
\begin{pmatrix}
J & \lambda &0 \\
K & -\mu ^{\prime}& 0 
\end{pmatrix}  \times \langle \theta _{n},K| \hat{Q}_{\lambda \mu ^{\prime}}| 0_{\text{def}}\rangle .
\end{eqnarray*}
\end{widetext}

\section{Symmetry considerations for cartesian transition densities constructed from axial QRPA calculations}
\label{Symmetry_considerations}

In this section, we show how to obtain the QRPA intrinsic transition density in the Cartesian coordinates, by  transforming the results obtained in cylindrical space ($r_{\perp},z$) with Eq.~(\ref{td}). The transition density at negative values of the Cartesian space,  can be obtained with the help of parity operator $\hat{\Pi}$ rules. The symmetry under parity inversion is commonly referred to as mirror symmetry, since it relates the function value at given coordinates to the value of the function evaluated at coordinates with opposite sign. Formally, the parity operator extracts such properties with changes in the coordinate sign according to $\hat{\Pi} \, |x\rangle = |-x \rangle $. Its eigenvalues $\hat{\Pi} | \pi\rangle = \pi |\pi \rangle $ are simply $\pi = \pm 1$, where $\pi = + 1$ defines positive (even) and $\pi = - 1$ negative (odd) parity functions. 
The action of the parity operator on the axially symmetric state brings in another phase due to the azimuthal angle, i.e. $\hat{\Pi}\, e^{iK\phi} = (-1)^{K}\, e^{iK\phi}$. Therefore, when reconstructing the intrinsic transition density in the Cartesian coordinates, we will make use  of these symmetry considerations and keep in mind that the intrinsic transition densities are obtained for a well-defined $K^{\pi}$.

\begin{figure}[htbp]
\centering
  \includegraphics[scale=0.5]{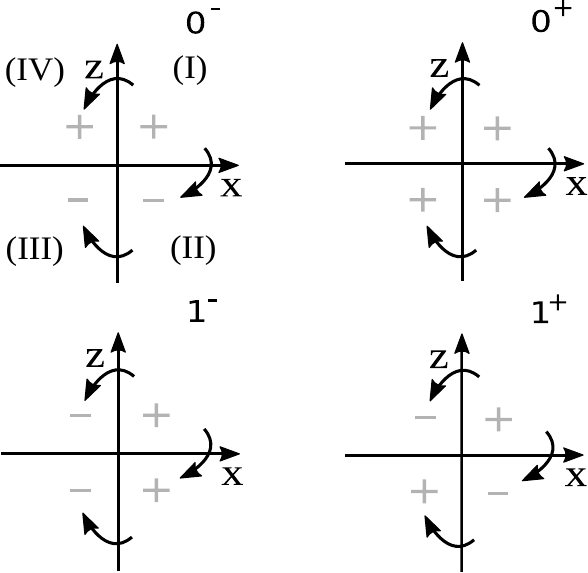}
\caption{\label{td-symmetry} Schematic representation of sign change rules for $K^{\pi} =0^{\pm}$ and $K^{\pi} =1^{\pm}$ cases when building the transition density from cylindrical (I) to Cartesian coordinate system. One starts from (I) with a given sign on the top right quadrant and a clock-wise rotate until (III) can be performed by adjusting the sign properly. The quadrant (IV) can be easily obtained from (I). }
\end{figure}

Our cylindrical basis-based model, allows us to restrict the computation of intrinsic transition densities to the two-dimensional space defined by positive-valued coordinates in the cylindrical coordinate system $z>0$ and $r_{\perp}$. 
Using this information, and the fact that the transition density transforms like the QRPA excited state, we construct the intrinsic transition density in Cartesian coordinates.
To simplify the notation, we suppress the $K^{\pi}$ information notation from the transition density since it is implicitly given by the QRPA state.
We consider the four blocks of the two-dimensional subspace (in Cartesian coordinates) separately:
\begin{itemize}
\item[]Block (I):  $x>0$ and $z>0$ ,  
\item[]Block (II): $x>0$ and $z<0$ ,  
\item[]Block (III): $x<0$ and $z<0$ ,  
\item[]Block (IV): $x<0$ and $z>0$  .
\end{itemize}
In the first Block (I), we simply have $\rho (x>0,z>0) = \rho (r_{\perp},z)$ with $x = r_{\perp}$.
For Block (II), we obtain $\rho (x>0,z<0) = (-1)^{K}\pi \, \rho (x>0,z>0)$, i.e. the transition density changes sign for unnatural parity states $(K^{\pi} = 0^{-},1^{+},\ldots)$ and is symmetric for natural-parity states $(K^{\pi} = 0^{+},1^{-},\ldots)$ \footnote{Alternatively, this part of the coordinate system can be directly obtained at the single particle level}. In Block (III), we take the real part of of the azimuthal-angle phase factor Re$[e^{iK\phi}]$ to construct the density with: $\rho (x<0,z<0) = (-1)^{K} \rho (x>0,z<0)$. Finally, in Block (IV), we have: $\rho (x<0,z>0) = (-1)^{K} \rho (r_{\perp},z>0)$ with $x = r_{\perp}$. The choice of using x or y coordinates here was merely for visualization purposes. The resulting phase changes are schematically illustrated in Figure (\ref{td-symmetry}) for $K^{\pi} =0^{\pm}$ and $K^{\pi} =1^{\pm}$, respectively.

\begin{figure}[htbp]
\centering
  \includegraphics[scale=.56]{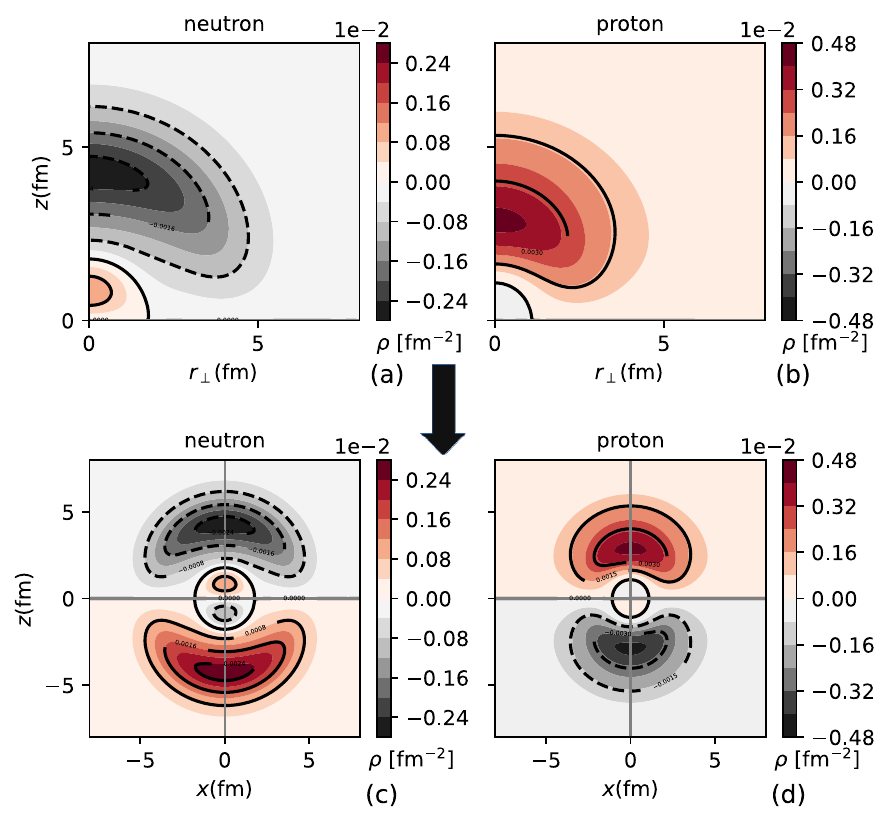}
\caption{\label{example_0m} QRPA intrinsic transition density of neutrons (left) and protons (right) for the $E = 17.22$ MeV GDR state of $^{96}$Zr with $K^{\pi}=0^{-}$. Panels (a,b) present densities for positive-valued cylindrical coordinates, while (c,d) show the x-z plane of the Cartesian space.}
\end{figure}

\begin{figure}[htbp]
\centering
  \includegraphics[scale=.56]{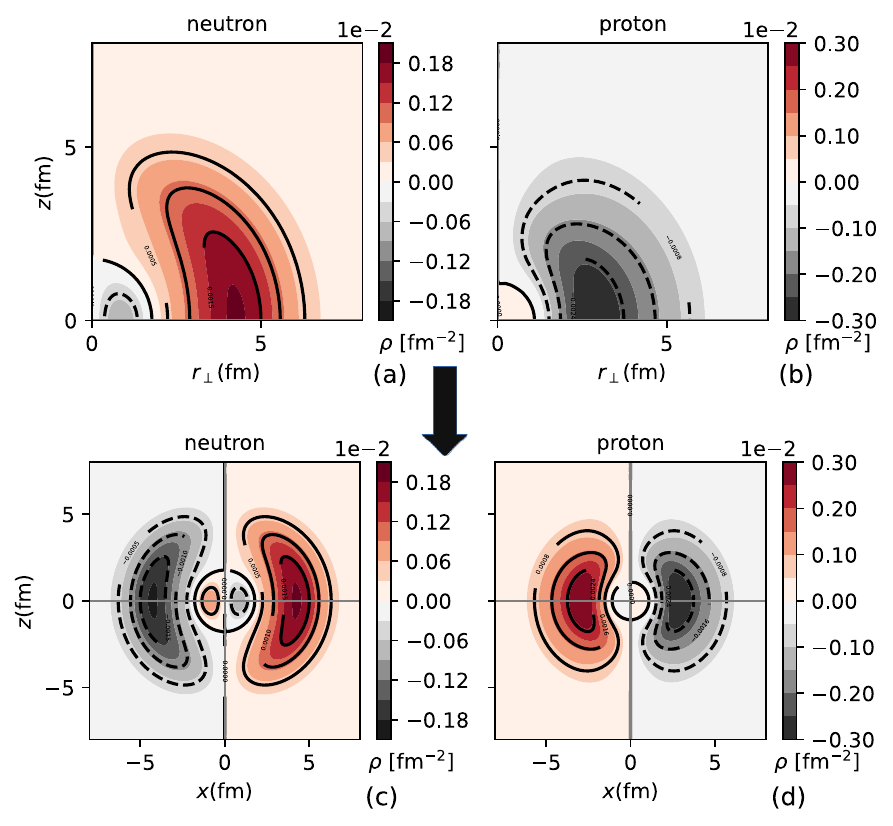}
\caption{\label{example_1m}QRPA intrinsic transition density of neutrons (left) and protons (right) for the $E = 17.22$ MeV GDR state of $^{96}$Zr with $K^{\pi}=1^{-}$. Panels (a,b) present densities for positive-valued cylindrical coordinates, while (c,d) show the x-z plane of the Cartesian space.}
\end{figure}

Figures (\ref{example_0m}-\ref{example_1m}) show proton and neutron intrinsic transition densities for the $K^{\pi}=0^{-}$ and $K^{\pi}=1^{-}$ components associated with GDR states in $^{96}$Zr. The top panels depict densities in the cylindrical coordinate system, while the bottom panels give the transition densities in Cartesian coordinates, using the block structure discussed above. The transformations shown here can be applied to any $K^{\pi}$ states within our formalism.

\clearpage
\begin{acknowledgments}
This work is performed in part under the auspices of the U.S. Department of Energy by Lawrence Livermore National Laboratory under Contract DE-AC52-07NA27344 with support from LDRD project 19-ERD-017. The work at Brookhaven National Laboratory was sponsored by the Office of Nuclear Physics, Office of Science of the U.S. Department of Energy under Contract No. DE-AC02-98CH10886 with Brookhaven Science Associates, LLC. We appreciate important contributions to the development of integrated nuclear structure and reaction theory by the late Eric Bauge - he played a central role in establishing the present CEA-LLNL collaboration. 
\end{acknowledgments}


\bibliography{ref.bib}

\end{document}